\documentclass[twocolumn,superscriptaddress,floatfix,preprintnumbers]{revtex4-1}
\usepackage{graphics,amssymb,amsmath,color}
\usepackage{graphicx}
\usepackage{bm}
\usepackage{hyperref}
\begin{document}

\title{Generation and Detection of Atomic Spin Entanglement in Optical Lattices}

%\author{Authors}
\author{Han-Ning Dai}
\affiliation{Physikalisches Institut, Ruprecht-Karls-Universit\"{a}t Heidelberg, Im Neuenheimer Feld 226, 69120 Heidelberg, Germany}
\affiliation{Hefei National Laboratory for Physical Sciences at Microscale and Department of Modern Physics, University of Science and Technology of China, Hefei, Anhui 230026, China.}
\affiliation{CAS Centre for Excellence and Synergetic Innovation Centre in Quantum Information and Quantum Physics, University of Science and Technology of China, Hefei, Anhui 230026, China}
\author{Bing Yang}
\affiliation{Physikalisches Institut, Ruprecht-Karls-Universit\"{a}t Heidelberg, Im Neuenheimer Feld 226, 69120 Heidelberg, Germany}
\affiliation{Hefei National Laboratory for Physical Sciences at Microscale and Department of Modern Physics, University of Science and Technology of China, Hefei, Anhui 230026, China.}
\affiliation{CAS Centre for Excellence and Synergetic Innovation Centre in Quantum Information and Quantum Physics, University of Science and Technology of China, Hefei, Anhui 230026, China}
\author{Andreas Reingruber}
\affiliation{Physikalisches Institut, Ruprecht-Karls-Universit\"{a}t Heidelberg, Im Neuenheimer Feld 226, 69120 Heidelberg, Germany}
\affiliation{Department of Physics and Research Center OPTIMAS, University of Kaiserslautern, Erwin-Schroedinger-Strasse, Building 46, 67663 Kaiserslautern, Germany}
\author{Xiao-Fan Xu}
\affiliation{Physikalisches Institut, Ruprecht-Karls-Universit\"{a}t Heidelberg, Im Neuenheimer Feld 226, 69120 Heidelberg, Germany}
\author{Xiao Jiang}
\affiliation{Hefei National Laboratory for Physical Sciences at Microscale and Department of Modern Physics, University of Science and Technology of China, Hefei, Anhui 230026, China.}
\affiliation{CAS Centre for Excellence and Synergetic Innovation Centre in Quantum Information and Quantum Physics, University of Science and Technology of China, Hefei, Anhui 230026, China}
\author{Yu-Ao Chen}
\affiliation{Hefei National Laboratory for Physical Sciences at Microscale and Department of Modern Physics, University of Science and Technology of China, Hefei, Anhui 230026, China.}
\affiliation{CAS Centre for Excellence and Synergetic Innovation Centre in Quantum Information and Quantum Physics, University of Science and Technology of China, Hefei, Anhui 230026, China}
\author{Zhen-Sheng Yuan}
\affiliation{Hefei National Laboratory for Physical Sciences at Microscale and Department of Modern Physics, University of Science and Technology of China, Hefei, Anhui 230026, China.}
\affiliation{CAS Centre for Excellence and Synergetic Innovation Centre in Quantum Information and Quantum Physics, University of Science and Technology of China, Hefei, Anhui 230026, China}
\affiliation{Physikalisches Institut, Ruprecht-Karls-Universit\"{a}t Heidelberg, Im Neuenheimer Feld 226, 69120 Heidelberg, Germany}
\author{Jian-Wei Pan}
\affiliation{Hefei National Laboratory for Physical Sciences at Microscale and Department of Modern Physics, University of Science and Technology of China, Hefei, Anhui 230026, China.}
\affiliation{CAS Centre for Excellence and Synergetic Innovation Centre in Quantum Information and Quantum Physics, University of Science and Technology of China, Hefei, Anhui 230026, China}
\affiliation{Physikalisches Institut, Ruprecht-Karls-Universit\"{a}t Heidelberg, Im Neuenheimer Feld 226, 69120 Heidelberg, Germany}

%email[]{Your e-mail address}
%\homepage[]{Your web page}
%\thanks{}
%\altaffiliation{}
%\affiliation{Physikalisches Institut, Ruprecht-Karls-Universit\"{a}t Heidelberg, Im Neuenheimer Feld 226, 69120 Heidelberg, Germany}
%\affiliation{Hefei National Laboratory for Physical Sciences at Microscale and Department of Modern Physics, University of Science and Technology of China, Hefei, Anhui 230026, China.}
%\affiliation{CAS Centre for Excellence and Synergetic Innovation Centre in Quantum Information and Quantum Physics, University of Science and Technology of China, Hefei, Anhui 230026, China}
%\affiliation{Department of Physics and Research Center OPTIMAS, University of Kaiserslautern, Erwin-Schroedinger-Strasse, Building 46, 67663 Kaiserslautern, Germany}

%Collaboration name if desired (requires use of superscriptaddress
%option in \documentclass). \noaffiliation is required (may also be
%used with the \author command).
%\collaboration can be followed by \email, \homepage, \thanks as well.
%\collaboration{}
%\noaffiliation

\date{\today}

\begin{abstract}
% insert abstract here
Ultracold atoms in optical lattices offer a great promise to generate entangled states for scalable quantum information processing owing to the inherited long coherence time and controllability over a large number of particles. We report on the generation, manipulation and detection of atomic spin entanglement in an optical superlattice. Employing a spin-dependent superlattice, atomic spins in the left or right sites can be individually addressed and coherently manipulated by microwave pulses with near unitary fidelities. Spin entanglement of the two atoms in the double wells of the superlattice is generated via dynamical evolution governed by spin superexchange. By observing collisional atom loss with in-situ absorption imaging we measure spin correlations of atoms inside the double wells and obtain the lower boundary of entanglement fidelity as \hbox{$\textrm{0.79}\!\pm\!\textrm{0.06}$}, and the violation of a Bell's inequality with \hbox{$\textrm{\it S}\!=\!\textrm{2.21}\!\pm\!\textrm{0.08}$}. The above results represent an essential step towards scalable quantum computation with ultracold atoms in optical lattices.
\end{abstract}

% insert suggested PACS numbers in braces on next line
%\pacs{}
% insert suggested keywords - APS authors don't need to do this
%\keywords{}

%\maketitle must follow title, authors, abstract, \pacs, and \keywords
\maketitle

% body of paper here - Use proper section commands
% References should be done using the \cite, \ref, and \label commands

During the last decades, quantum entanglement, the key resource for quantum information processing\cite{nielsen2010quantum}, has been created in many systems like photons\cite{pan2012multiphoton}, ions\cite{blatt2008entangled}, superconducting circuits\cite{devoret2013superconducting}, and solid-state qubits\cite{awschalom2013quantum}. They have been wildly used for studying quantum computation\cite{ladd2010quantum} and quantum simulation\cite{trabesinger2012quantum}. Nowadays, a significant demand towards scalable quantum information processing is to efficiently construct multipartite entangled states.
Since ultracold atoms in an optical lattice\cite{bloch2008quantum} have excellent coherence properties and can be manipulated in parallel, an attractive protocol\cite{vaucher2008creation} was proposed to create resilient entangled states for measurement-based quantum computation\cite{raussendorf2001one,walther2005experimental} with optical superlattices. In this protocol, maximally entangled Bell-type states are first prepared in double well arrays (DWs) of a superlattice, then these Bell pairs are connected to each other to create cluster states by using Ising-type superexchange interactions\cite{duan2003controlling}, and finally a computational algorithm is implemented by performing single-particle measurements together with unitary operations.

Along this direction, a preliminary step has been taken in an exciting experiment by observing and controlling the superexchange interactions between two atomic spins in optical superlattices\cite{trotzky2008time}. However, due to difficulties for addressing single spins and measuring spin correlations in superlattices, it remains a challenge to generate, characterize, and manipulate the entanglement of atomic spins, which are the essential ingredients for measurement based quantum computation\cite{vaucher2008creation,bloch2008quantum}.

In this work, we developed a new superlattice configuration featured with spin dependence, which allows us to individually address and manipulate the spins in the left or right sites of the DWs with microwave (MW) pulses. Relying on this configuration, high-fidelity initialization of the DWs from \hbox{$|\!\downarrow,\downarrow\rangle$} to \hbox{$|\!\uparrow,\downarrow\rangle$} is achieved (\hbox{$|\!\uparrow\rangle$}  and \hbox{$|\!\downarrow\rangle$ } denote the two spin states while the comma separates the left and right occupations). The entangled state, \hbox{$ (|\!\uparrow,\downarrow\rangle+i|\!\downarrow,\uparrow\rangle)/\sqrt{2}$}, is generated via a $\sqrt{\textrm{SWAP}}$  operation\cite{anderlini2007controlled,bloch2008quantum}. Afterwards, an effective short-range gradient magnetic field is induced to manipulate the phase of the entangled state and transfer it to the Bell state \hbox{$(|\!\uparrow,\downarrow\rangle+|\!\downarrow,\uparrow\rangle)/\sqrt{2}$}. Then, employing a new detection routine with two-stage filtering and imaging, we measure quantitatively the spin correlations, \hbox{$\langle\hat{S}_z\!\otimes\!\hat{S}_z\rangle$} and \hbox{$\langle\hat{S}_y\!\otimes\!\hat{S}_y\rangle$} of the two atoms by observing the atom loss arising from hyperfine changing collisions\cite{julienne1997collisional,schmaljohann2004dynamics}. The lower boundary of the entanglement fidelity and the violation of the Clauser-Horne-Shimony-Holt (CHSH) Bell inequality\cite{clauser1969proposed} are finally derived from the spin correlations.

\begin{figure*}[!t]
\centerline{\includegraphics[width=12.13 cm]{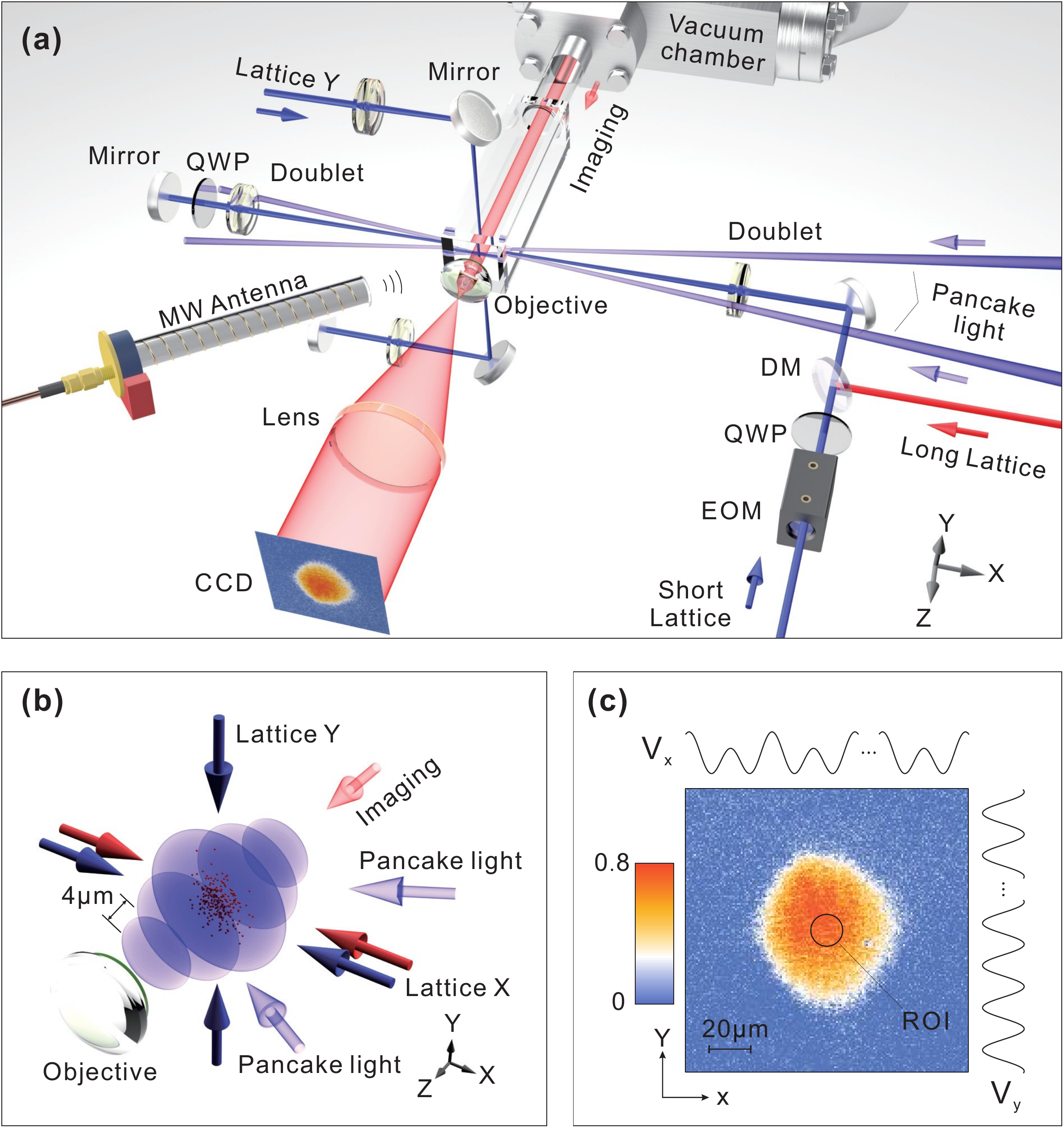}}%
\caption{{\bf The experimental apparatus.} \textbf{(a)} Three retro-reflected optical lattices create the double well arrays: a short lattice $V_{xs}$, a long lattice $V_{xl}$ along the X direction, and lattice $V_y$ along the Y direction. The imaging beam is applied along the Z direction. \textbf{(b)} The laser beams around the atom cloud. The 2D gas is loaded into a single layer of a pancake-shaped trap, which is a series of 4$\mu$m-distance layers created by interfering two beams. \textbf{(c)} The density distribution of the MI derived from an average over 10 samples by in-situ absorption imaging. The central part of the atom cloud with a diameter of 16 pixels (0.93 $\mu$m per pixel) is used as the region of interest (ROI) for later studying spin dynamics and entanglement. The averaged filling in the ROI is about 0.8 atoms per site. The geometries of the lattices are shown on the X and Y direction.}
\end{figure*}

\begin{figure*}[!t]
\centerline{\includegraphics[width=13.593 cm]{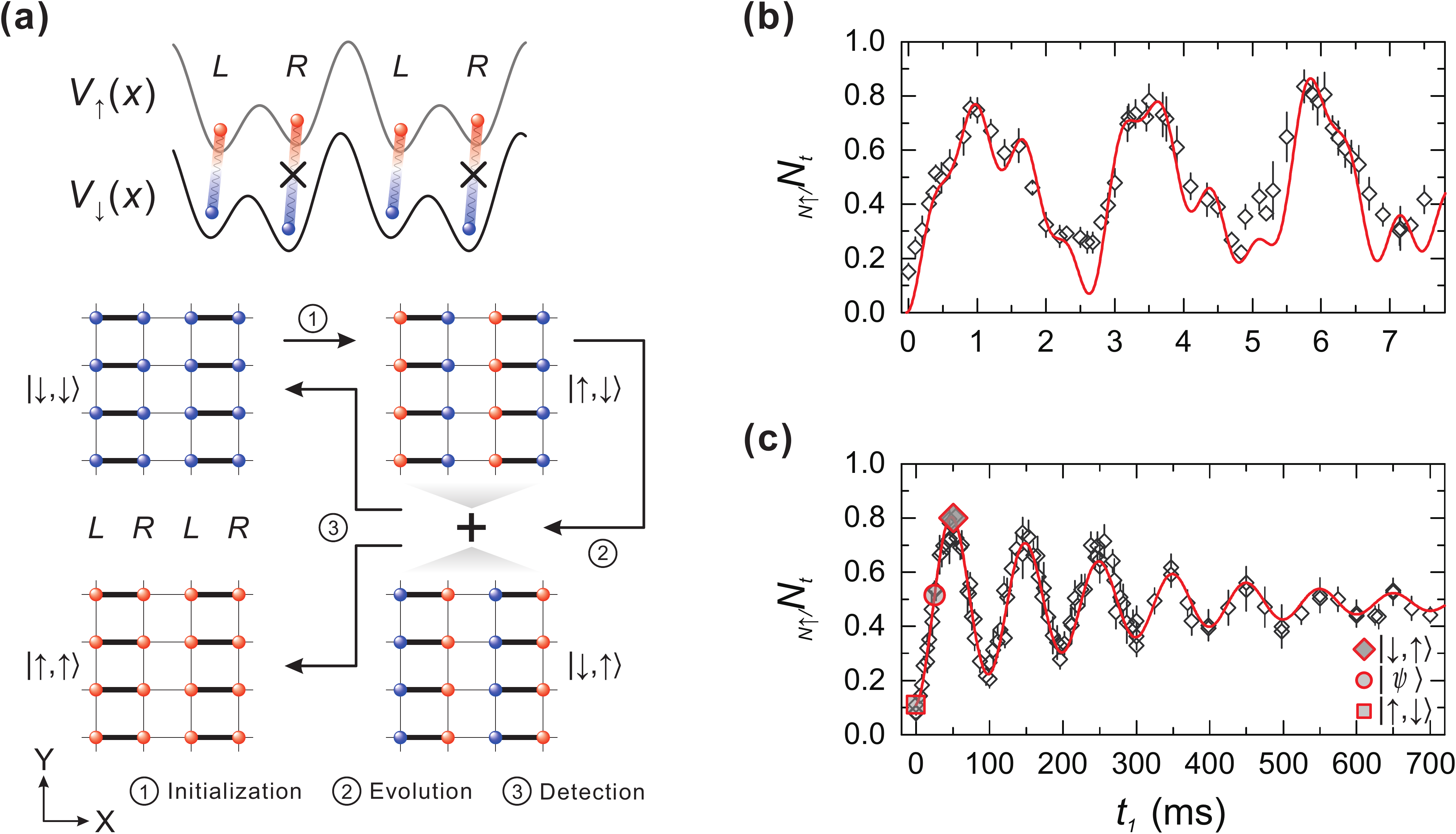}}
\caption{{\bf Observation of spin dynamics driven by the superexchange interaction.} \textbf{(a)} Initialization, evolution and detection of spin states in the spin-dependent superlattice. The lattice potentials for  $|\!\downarrow\rangle$  and $|\!\uparrow\rangle$  are different, and the coupling frequency of \hbox{$|\!\downarrow\rangle \Leftrightarrow |\!\uparrow\rangle$ } in the left site is shifted 31.8 kHz away from that in the right site. The spins in left sites can be individually addressed with  a MW pulse, and the DW can be initialized from $|\!\downarrow,\downarrow\rangle$ to $|\!\uparrow,\downarrow\rangle$. Then the spin state is evolved to a superposition state by superexchange driven evolution. By flipping the spin in every left site before the imaging pulse, the DW states of $|\!\uparrow,\downarrow\rangle $ and $|\!\downarrow,\uparrow\rangle $ are transferred to $|\!\downarrow,\downarrow\rangle$ and $|\!\uparrow,\uparrow\rangle$, respectively, for detection. Only the $|\!\uparrow\rangle$ can be coupled with the imaging pulse. With this detection method, the superexchange driven evolution is observed by recording the population of $|\!\uparrow\rangle$ in two experimental conditions: \textbf{(b)} $V_{xs}=$12 Er, $V_{xl}=$10 Er, $J/U = 0.37$, a numerical simulation of the BHM agrees well with the experiment data; \textbf{(c)} $V_{xs}=$20 Er, $V_{xl}=$10 Er, $J/U = 0.04$, a damped sine fit shows a superexchange frequency of 10.0(5) Hz with a $1/e$ lifetime of 280(20) ms.}
\end{figure*}

Here we stress that the atom cloud is prepared in a pancake-shaped trap deeply in 2D regime on the X-Y plane, therefore the longitudinal inhomogeneity along the Z direction existing in 3D lattices is excluded.
Moreover, an in-situ absorption imaging along the Z direction is employed for detecting the density distribution of the atoms,  which has a great advantage of suppressing the transversal inhomogeneity (in X-Y plane) by selecting a proper area in the sample. The two approaches are essential for generating entangled states with high fidelity and long coherence time. The present techniques and experimental routines form an architecture for generating scalable cluster states\cite{vaucher2008creation} as well as many-body Hamiltonians\cite{bloch2008many}, therefore are ideally suited for studying measurement-based quantum computation\cite{raussendorf2001one} and performing quantum simulation\cite{lewenstein2012ultracold}.

%\section*{Scheme for generating spin entanglement in the superlattice}
The system under consideration is a series of isolated double-well arrays created by an optical superlattice --- a system which can be well described with the Bose-Hubbard model characterized by a nearest-neighbor tunneling $J$, onsite interaction energy $U$ and the imbalance offset $\Delta$ in DWs\cite{duan2003controlling}. All these parameters can be well controlled by the intensities and frequencies of the lattice lasers. In the following part we focus on the condition of balanced DWs ($\Delta\!=\!0$) and deep lattices where the interaction dominates $U\!\gg \!J$. With the initial states in the subspace of singly filled states of \hbox{$|\!\uparrow,\downarrow\rangle$} and \hbox{$|\!\downarrow,\uparrow\rangle$}, double occupation at each site is allowed only virtually. Then the Bose-Hubbard model can be described as a Heisenberg spin model \hbox{$\hat{\bf H}\!=\!- J_{ex}  \hat{\bf S}_{L}\!\cdot\!\hat{\bf S}_{R} $}, where \hbox{$J_{ex}\!=\!4J^2/U$} denotes the superexchange coupling between the two sites in a DW. By initializing the DW to \hbox{$|\!\uparrow,\downarrow\rangle$} and using appropriate potentials, one can observe the superexchange driven evolution from \hbox{$|\!\uparrow,\downarrow\rangle$} to \hbox{$|\!\downarrow,\uparrow\rangle$}, and vice versa\cite{trotzky2008time}. A $\sqrt{\textrm{SWAP}}$ operation is applied to generate the entangled state \hbox{$|\psi\rangle\!=\!(|\!\uparrow,\downarrow\rangle\!+\! i|\!\downarrow,\uparrow\rangle)/\sqrt{2}$} by halting the superexchange driven evolution at \hbox{$t_{1} \!=\! h/4J_{ex}$}, with $h$ being the plank constant. However, we intend to prepare the Bell state \hbox{$|t\rangle \!=\! (|\!\uparrow,\downarrow\rangle\!+\!|\!\downarrow, \uparrow\rangle) / \sqrt{2}$} (the spin triplet), which is taken as the starting point for creating cluster states for measurement based quantum computation\cite{vaucher2008creation}. In order to achieve $|t\rangle$ from $|\psi\rangle$, an additional phase between the two components \hbox{$|\!\uparrow,\downarrow\rangle$} and \hbox{$|\!\downarrow,\uparrow\rangle$} has to be introduced. We first induce an effective gradient magnetic field inside the double-well by creating a spin-dependent\cite{anderlini2007controlled} superlattice. This causes non-degeneracy between the two components with an energy split of $\delta$. By holding the atoms in this lattice for a period $t_2$, an additional phase of \hbox{$\delta\!\cdot \!t_2$} is accumulated. Therefore the phase of the entangled state can be modulated and the spin triplet $|t\rangle$ is obtained by controlling $t_2$.

%\section*{Experimental setup}
Our experiment starts from preparing a two dimensional (2D) quantum gas by loading a nearly pure $^{87}$Rb Bose-Einstein condensate (BEC) into a single layer of a pancake-shaped trap, which is created by interfering two laser beams with wavelength of $\lambda_s=$767 nm and an angle of intersection of 11$^\circ$, as shown in Fig.1(a,b). The confinement of this trap is highly anisotropic, with trap frequencies $\omega_{x,y,z}\!\approx\!2\pi \times $(16,14,7000) Hz, leading to an aspect ratio of 467:1. This 2D sample contains around 1.2$\times10^4$ atoms spin-polarized in $|\!\downarrow\rangle$ and reaches a temperature as low as $T_{\rm 2D}=23(2)$ nK. We use the hyperfine levels $|\!\downarrow\rangle\!=\!|F=1,m_F=-1\rangle$ and $|\!\uparrow\rangle\!=\!|F=2, m_F=-2\rangle$ to represent the pseudo spin-1/2 system. The ratio $\hbar\omega_{z}/k_{\rm B}T_{\rm 2D}\!\approx\!14.6$ with $k_{\rm B}$ being the Boltzmann constant, indicates that the sample is deeply in the 2D regime\cite{dalfovo1999theory}. %A distinct feature of ultracold atoms in a single 2D layer is the suppressed longitudinal inhomogeneity (along the Z direction), which exists in 3D systems with multi layers.

Then the atoms are adiabatically loaded into a square lattice, which is composed of two retro-reflected lattices $V_{xs}$ (short lattice)  and $V_y$ (Y lattice) along the X and Y directions with wavelength $\lambda_{xs}\!=\!\lambda_{y}\!=\!767$ nm, as shown in Fig.1(a,b). The system enters the Mott insulator (MI) regime\cite{jaksch1998cold,greiner2002quantum,jordens2008mott,schneider2008metallic} where most sites are filled with single atoms when the lattices are ramped to $V_{xs}\!=\!V_y\!=\!25$ Er, with $\textrm{Er}\!=\!h^2/2m\lambda_{xs}^2$ being the recoil energy and $m$ the mass of the atom. Afterwards the two lattices are increased further to $V_{xs}\!=\!60$ Er and $V_{y}\!=\!40$ Er to freeze out the atom tunneling. The density distribution of the sample is obtained by in-situ absorption imaging along the Z direction with a microscope objective ($\textrm{N.A.}\!=\!0.48$) and a low noise CCD camera, shown in Fig.1(c). An average filling number of 0.80 atoms per site in the center part of the atom cloud is obtained. Measuring  the atom loss of hyperfine changing collisions in the lattices reveals that 76\% of the lattice sites are filled with one atom, about 2\% are filled with two atoms and the rest are vacant sites\cite{EntangleSuperlatticeSI}. A great advantage of using in-situ imaging is that one can select a proper area in the sample with optimal filling properties and least transversal inhomogeneity (see later text).

Next, the MI atoms are transferred into the DWs of a superlattice along the X direction. By superimposing another retro-reflected lattice $V_{xl}$ (long lattice, wavelength $\lambda_{xl}\!=\!1534$ nm) with the short lattice, a superlattice \hbox{$V_x(x)\!=\!V_{xs}\cos^2(k_x x)\!-\!V_{xl}\cos^2(k_x x/2 +\varphi)$} is formed, where $k_x\!=\!2\pi/\lambda_{xs}$  is the wave vector and $\varphi$ is the superlattice phase. Controlling the relative frequency between the long and short lattices, the superlattice phase is tuned to 0, therefore balanced DWs with spin state $|\!\downarrow,\downarrow\rangle$ are prepared.

%\section*{Superexchange driven dynamics and transversal inhomogeneity}
\begin{figure*}[!t]
\centerline{\includegraphics[width=14.653 cm]{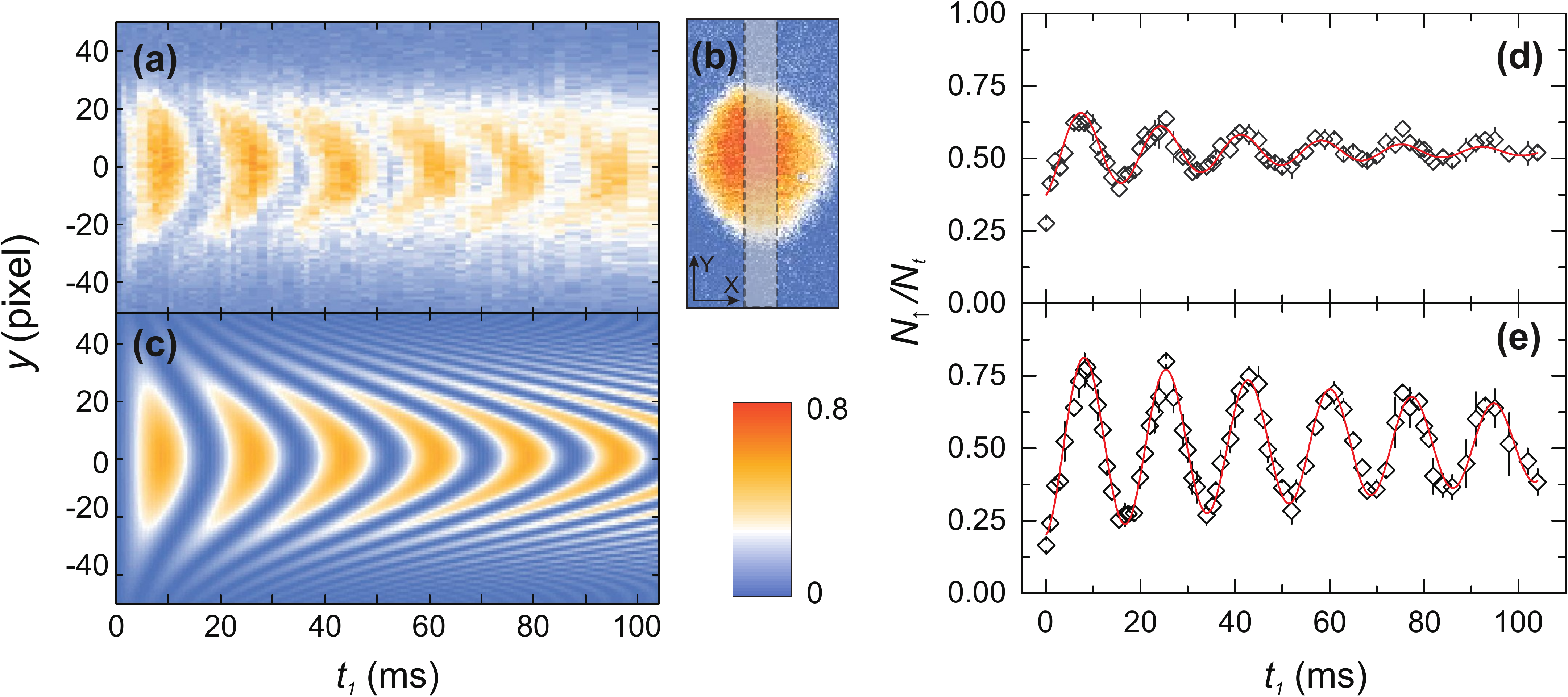}}
\caption{{\bf Transversal inhomogeneity observed in superexchange evolution.} \textbf{(a)} Experimental time evolution of line density profiles during the superexchange dynamics in the condition of $V_{xl} = 10$ Er and $V_{xs} = 16$ Er (\hbox{$J/U\!=\!0.11$}). \textbf{(b)} Each line density profile is the mean cross section along the Y direction by averaging the rectangular region (14.9 $\mu$m width) along the X direction. A dephasing process is clearly visible along the Y direction.
\textbf{(c)} The simulation of experiment with the measured laser diameters $w_{xs}\!=\!240$ $\mu$m, $w_{xl}\!=\!300$ $\mu$m and $w_y\!= \!240$ $\mu$m. The evolution of the superexchange resolved by using either the whole atom cloud \textbf{(d)}  or using the center part (diameter of 14.9 $\mu$m) \textbf{(e)} as the ROI for counting atoms. In the first case, a damped sine fit gives a $1/e$ life time of 41(5) ms, while 115(12) ms for the second one.}
\end{figure*}
We then initialize all the DWs to $|\!\uparrow,\downarrow\rangle$ by flipping the atomic spins in every left site inside the spin-dependent superlattice with a MW pulse. Concretely, tuning the voltage applied to the electro-optical modulator (EOM), an angle of $\theta$ between the polarizations of the incident and the retro-reflected beams of the short lattice is created, therefore a spin-dependent superlattice is built up\cite{EntangleSuperlatticeSI}. In the condition of $V_{xl}\!=\!56.3$ Er, $V_{xs}\!=\!150$ Er and $\theta \!= \!46^{\circ}$, the resonant microwave frequency for coupling $|\!\downarrow\rangle_L\Leftrightarrow|\!\uparrow\rangle_L$ in the left site is shifted 31.8 kHz away from the resonance $|\!\downarrow\rangle_R\Leftrightarrow|\!\uparrow\rangle_R$  of the right site, shown in Fig.2(a). A high addressing fidelity is achieved by using a MW $\pi$-pulse with Rabi frequency of $\Omega=2\pi\times8.1$ kHz, during which the magnetic field $B_x =95$ $\mu$T along the X direction is actively stabilized and the noise is suppressed to less than 16 nT.

After state initialization, the spin dependence is switched off by ramping down the EOM voltage. Meanwhile the quantum axis is rotated from X to Z direction to avoid the residual magnetic gradient along the X direction. The superexchange is now initialized by first ramping down the long lattice to 10 Er, and afterwards ramping down the short lattice to the final value $V_{\rm SE}$ in 500 $\mu$s. After letting the system evolve for a time $t_1$, we halt the superexchange by ramping up the short lattice to 60 Er in 500 $\mu$s. The spin configuration is then freezed until the state detection.

To detect the spin state inside the DWs, the quantum axis is aligned to the X direction, then the spin-dependence is switched on for addressing the left sites. Flipping the spins in the left sites, the two spin configurations will end up along two different channels: $|\!\uparrow,\downarrow\rangle \Rightarrow |\!\downarrow,\downarrow\rangle$ and $|\!\downarrow,\uparrow\rangle \Rightarrow |\!\uparrow,\uparrow\rangle$, as shown in Fig.2(a). Only the $|\!\uparrow\rangle$ state can absorb the imaging light (cycling transition $|5S_{1/2},F=2\rangle \Leftrightarrow|5P_{3/2},F=3\rangle$). Therefore, the superexchange-driven oscillation between $|\!\uparrow,\downarrow\rangle$ and $|\!\downarrow,\uparrow\rangle$ can then be observed with the in-situ imaging by comparing the population of atoms in spin up $N_{\uparrow}$ to the total atom number in the ROI $N_t$. Two typical oscillations measured at different DW potentials are shown in Fig.2. For a low barrier height ($J/U=0.37$), the dynamics can be directly described with the Bose-Hubbard model, in which the fast oscillations correspond to the first-order tunnelings. The simulation matches well with the experiment as shown in Fig.2(b). For larger barrier heights ($J/U\ll1$), the oscillation is then mainly driven by the superexchange interaction. In the experimental condition of $V_{xl}\!=\!10$ Er, $V_{xs}\!=\! 20$ Er ($J/U = 0.04$), an oscillation frequency $10.0(5)$ Hz and a decay constant of 280(20) ms is obtained.

Here we study the dephasing effect of the superexchange evolution. As shown in Fig.3(a,b), the dephasing along the Y direction is clearly observed when we perform the superexchange in the DWs with \hbox{$V_{xl}\!=\ 10$} Er and \hbox{$V_{xs}\!=\!16$} Er. A simulation of the Bose-Hubbard model with considering the lattice profiles is shown in  Fig.3(c), which agrees well with the experimental data. We then find that the dephasing process is mainly caused by the transversal inhomogeneity of the lattice profiles through spatially modulating the Bose-Hubbard parameters of the DWs.
By using the in-situ imaging system, one can resolve the atom distribution with a resolution around 2 $\mu$m. Thus we can choose an area in the atom cloud as the ROI with least inhomogeneity to obtain a long coherence time while keep a good signal to noise ratio. Here, an area with diameter 14.9 $\mu$m in the center of the atom cloud is used as the ROI. The resulting coherence time of the superexchange is 3 times longer when using the ROI instead of the whole atom cloud, as shown Fig.3(d,e). The heating effect due to scattering the 767-nm lights of the pancake and lattices is considered as another limitation of the coherence.

%\section*{Generation of Bell states in the superlattice}
The $\sqrt{\textrm{SWAP}}$ operation is realized by halting the superexchange-driven evolution ($V_{xl} \!= \!10$ Er, $V_{xs} \!=\! 20$ Er) at $t_1\!=\!25.1$ ms, and the DWs are then prepared in the entangled state $|\psi\rangle$. In order to control the phase between the two components in this entangled state, we induce an effective short-range gradient magnetic field inside the DW by switching on the spin dependence with \hbox{$V_{xl}\!=\!5.6$} Er,  \hbox{$V_{xs}\!=\!60$} Er and \hbox{$\theta\!=\!7.5^\circ$}. This gradient creates a non-degeneracy of \hbox{$\delta\!=\!2\pi\!\times\!427$} Hz between the two components and initiates an oscillation\cite{trotzky2010controlling} between the spin singlet state (\hbox{$|s\rangle\!=\!(|\!\uparrow,\downarrow\rangle-|\!\downarrow,\uparrow\rangle)/\sqrt{2}$}) and the triplet state $|t\rangle$.  After a singlet-triplet oscillation (STO) time $t_2$, we switch off the spin dependence and ramp down the long lattice to 0.  By measuring in the rotated basis \hbox{$|\pm\rangle=(|\!\uparrow\rangle\pm i|\!\downarrow\rangle)/\sqrt{2}$}, i.e. applying a $\pi/2$-pulse on both DW sites, the triplet and the singlet state can be discriminated: \hbox{$|t\rangle \stackrel{\pi/2}{\Longrightarrow} (|\!\uparrow,\uparrow\rangle + |\!\downarrow,\downarrow\rangle)/\sqrt{2}$ }and \hbox{$|s\rangle \stackrel{\pi/2}{\Longrightarrow} |s\rangle$}.

\begin{figure}[!tb]
\centerline{\includegraphics[width=0.75\linewidth]{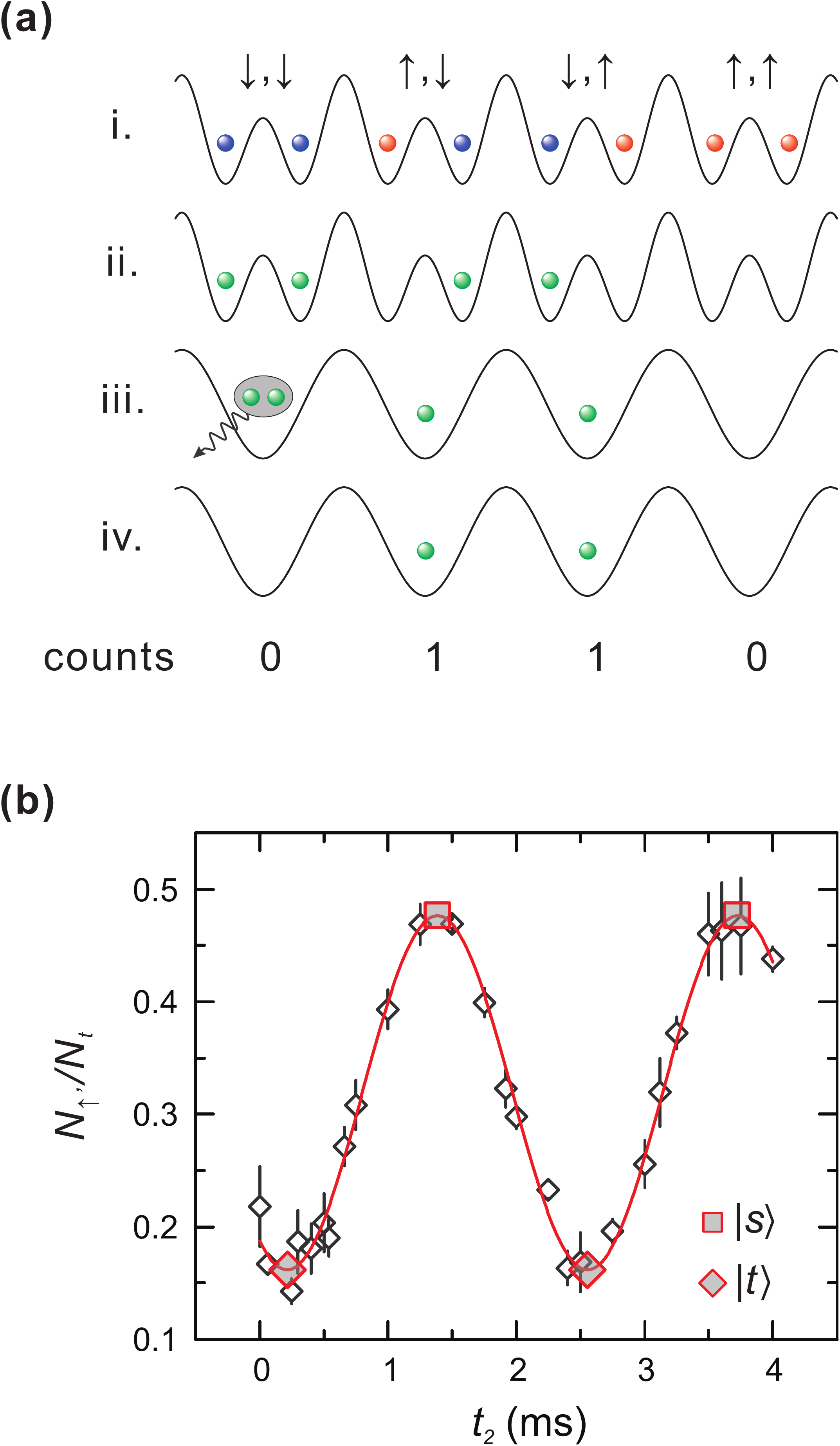}}%7.263 cm
\caption{{\bf Measurement of the entanglement phase.} \textbf{(a)} Two-stage filtering and imaging routine for detecting the phase of entanglement:
i) the complete spin correlated basis in the DW;
ii) remaining $|\!\uparrow^{\prime}\rangle$ atoms after recording the number of $|\!\uparrow\rangle$  atoms and the MW rapid adiabatic passage;
iii) merging the DWs and holding the atoms for 500ms in a deep lattice to remove the double occupancies;
iv) counting the number of remaining atoms $N_2$.
The four different initial pair states will contribute differently to the final atom counting, thus by projecting the entangled state to $|\pm\rangle$ basis one can resolve the singlet and triplet state.
\textbf{(b)} Time evolution of the remaining atom number for different STO time after the two-stage filtering and imaging process. A period of 2.34(2) ms is derived from a sine fit.}
\end{figure}

Then a two-stage filtering and imaging routine is employed to determine the entanglement phase, as shown in Fig.4(a). We assume the complete spin correlated basis in the DW, \{$|\!\downarrow,\downarrow\rangle$, $|\!\uparrow,\downarrow\rangle$, $|\!\downarrow,\uparrow\rangle$, $|\!\uparrow,\uparrow\rangle$\}. First, an imaging pulse is used to count all the spin-up atoms $N_1$ and remove them from the lattices by heating. Afterwards, a MW rapid adiabatic passage is applied to flip all the remaining atoms to \hbox{$|\!\uparrow^{\prime}\rangle = |F=2,m_F =-1\rangle$}. Then, the atoms are transferred to the long lattice of  35 Er in 2 ms by ramping down the short lattice. Meanwhile the lattice depth along the Y direction is increased to 60 Er for enhancing the onsite interactions. With this condition, the double occupancies will be removed from the lattice after a holding time of 500 ms due to hyperfine changing collisions, while the single fillings will survive. Then the in-situ imaging is performed for counting the remaining number of atoms $N_2$. By this method, the triplet state can not contribute any count to the final number of atoms, while one atom of the singlet will be counted. As shown in Fig.4(b), an oscillation curve with a period of 2.34(2) ms is obtained by counting the number of atoms in the ROI at different STO times with this detection routine. According to the STO curve, the phase of the entangled state can be well controlled and the Bell state $|t\rangle$ can be prepared after a STO time $t_2 = 0.3$ ms.

%\section*{Detecting spin entanglement in the superlattice}
Following the two-stage filtering and imaging routine, we verify the spin-entanglement in the DWs by comparing the sum of two images $(N_1+N_2)$ to the total number of atoms $N_t$  in the ROI. The number of atoms in the correlated basis of $|\!\downarrow,\downarrow\rangle$ is then derived as \hbox{$N_{\downarrow,\downarrow} = N_t - (N_1+N_2)$}. By addressing and flipping the spins on the left site, right site or both sites before the detection process, we can project the states $|\!\uparrow,\downarrow\rangle$, $|\!\downarrow,\uparrow\rangle$, or $|\!\uparrow,\uparrow\rangle$ to the measurement state $|\!\downarrow,\downarrow\rangle$ and derive the number of atoms for $N_{\uparrow,\downarrow}$, $N_{\downarrow,\uparrow}$, or $N_{\uparrow,\uparrow}$, respectively. From these measured spin fractions, we can get coincidence-like probabilities $P_{a,b} = N_{a,b}/\sum_{u,v}N_{u,v}$ ($a,b,u,v = \uparrow$ or $\downarrow$) and the spin correlation of $\langle\hat{ S}_{ z}\otimes\hat{ S}_{ z}\rangle$ is \hbox{$E_{zz} = P_{\uparrow,\uparrow} - P_{\uparrow,\downarrow} - P_{\downarrow,\uparrow} + P_{\downarrow,\downarrow}$}.

\begin{figure*}[!t]
\centerline{\includegraphics[width=13.443 cm]{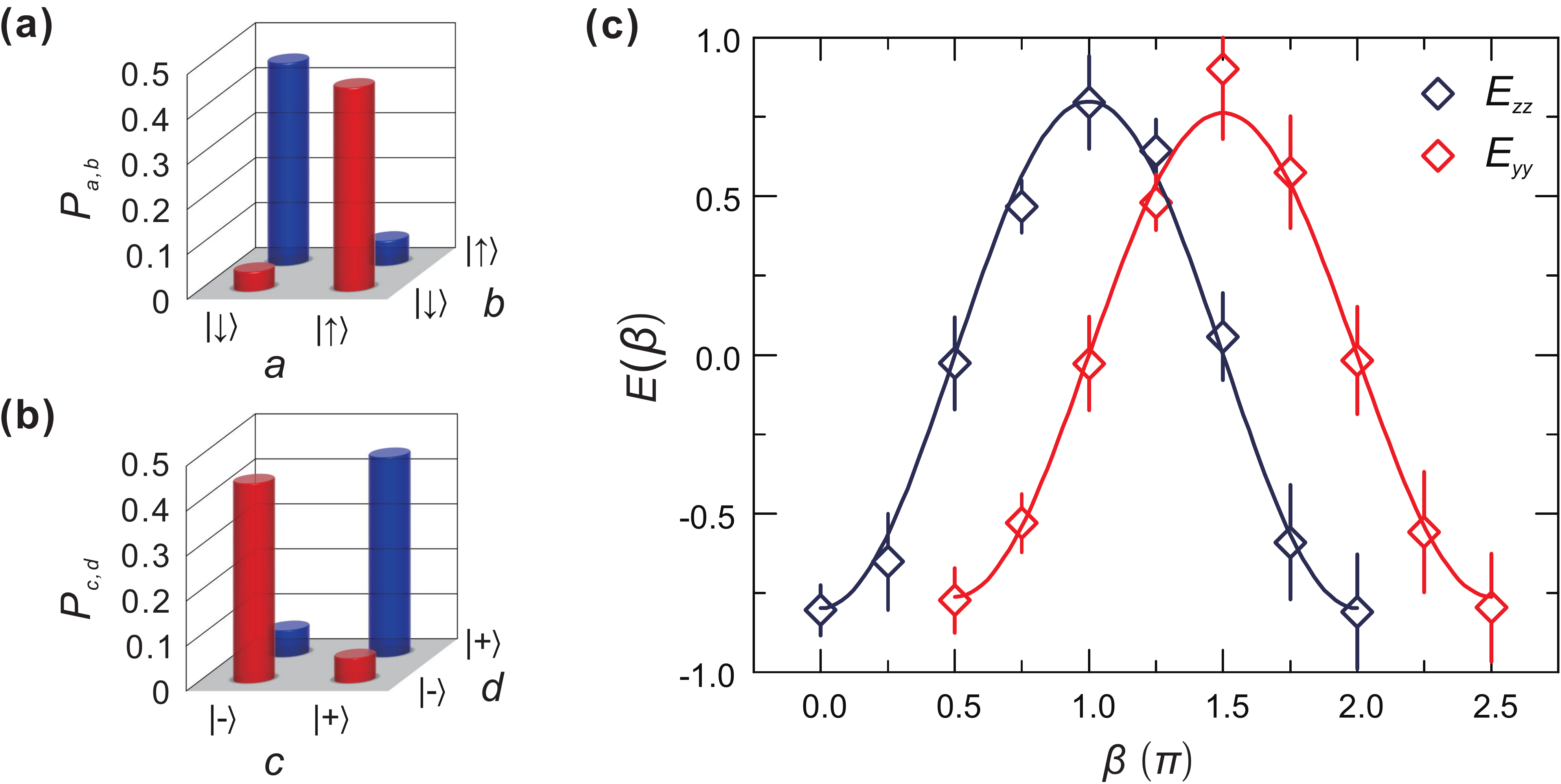}}
\caption{{\bf Measurement of spin correlations and entanglement.} \textbf{(a-b)} The coincidence-like probabilities derived in $|\!\uparrow\rangle/|\!\downarrow\rangle$ and $|+\rangle/|-\rangle$ basis, $P_{\uparrow,\uparrow}=0.05(3)$, $P_{\uparrow,\downarrow}=0.45(3)$, $P_{\downarrow,\uparrow}=0.45(3)$, and $P_{\downarrow,\downarrow}=0.04(3)$; $P_{+,+}=0.44(5)$, $P_{+,-}=0.06(4)$, $P_{-,+}=0.05(4)$, and $P_{-,-}=0.44(5)$. \textbf{(c)} Measured spin correlation curves of $E_{zz}(\beta)$ and $E_{yy}(\beta)$. We derive the violation of the CHSH type Bell inequality as $2.21\pm0.08$ from the sine fittings.}
\end{figure*}

This technique for measuring spin correlations is similar to the measurement of polarization correlations in photonic entanglement\cite{pan2012multiphoton}. Similarly, the spin fractions in the $|+\rangle/|-\rangle$ basis can be measured by applying a $\pi/2$-pulse to both sites before these measurements, and the probabilities $P_{c,d}$ ($c,d=+$ or $-$) are derived. The spin correlation of $\langle\hat{S}_{y}\otimes\hat{S}_{y}\rangle$ is $E_{yy} = -P_{+,+} + P_{+,-} + P_{-,+} - P_{-,-}$. From these measurements, as shown in Fig.5(a,b), we obtain the lower boundary of the entanglement fidelity\cite{blinov2004observation} \hbox{$F\geq -(E_{zz}+E_{yy})/2= 0.79\pm0.06$},  higher than the classical limit of 0.5 by 5 standard deviations.

Furthermore, we demonstrate the existence of spin entanglement by violating the CHSH-type Bell inequality. The spin-correlation curves are measured by first assigning the left site an additional rotation $\beta = \alpha\pi/4$ ($\alpha= 1,2,3...,8$), and then repeating the above measurements for the spin correlations of $E_{zz}(\beta)$ and $E_{yy}(\beta)$, as shown in Fig.5(c). Choosing the rotation of $\alpha\pi/4$ for the left site causes minimum response to the atoms in the right site. From fitting the two spin-correlation curves, the quantity $S=|E(\theta_1,\theta_2)+E(\theta_1,\theta_2^\prime)- E(\theta_{1}^\prime,\theta_2)+E(\theta_1^{\prime},\theta_2^\prime)|$, with $(\theta_1,\theta_1^\prime, \theta_2,\theta_2^\prime)=(0,\pi/2,3\pi/4,5\pi/4)$, is derived as $S=2.21\pm0.08$, which violates the CHSH inequality with 2.7 standard deviations.

The observed entanglement fidelity is limited by several experimental imperfections. First,  due to the coupling with the environment, the spin states experience decoherence during the period of generating the entangled states and another holding time of stabilizing the magnetic field for further microwave operations. This decoherence causes a degradation of the fidelity by about 17\% . Second, the errors of microwave operations decrease the fidelity by about 3\%. Therefore the expected fidelity of 80\% is consistent with the experimental measurement. Although the generation of spin entanglement in double wells is deterministic, extending to longer chains of spin entanglement suffers from lattice defects due to finite temperature. Further cooling\cite{bakr2011orbital} down the temperature of the atom cloud will be helpful for suppressing vacant sites and achieving unit filling in the lattices.

%\section*{Summary and outlook}
In summary, we have demonstrated the generation, manipulation and detection of atomic spin entanglement in an optical superlattice, the first step towards measurement based quantum computation. The spin dependence built-in superlattice brings great flexibility for addressing the single spins in the double-well arrays, which leads to high-fidelity state initialization and detection. Both the longitudinal and transversal inhomogeneities are well suppressed by confining the ultracold atoms in a 2D plane and employing in-situ imaging. Therefore, a long coherence time of the entangled state is achieved. The routine of two-stage filtering and imaging makes it possible to measure the spin correlations from which we can derive the lower boundary of the entanglement fidelity as $0.79\!\pm\!0.06$ and the violation of the Bell's inequality with $S=2.21\pm0.08$. This method may be used for characterizing large scale entangled states by investigating the entropic inequalities\cite{alves2004multipartite}.

By connecting the Bell pairs, one can extend the entanglement to a lattice chain, and further to a 2D plane with the presently developed techniques and experimental routines. Large 2D cluster states can be generated as universal resources for quantum computation\cite{vaucher2008creation,jiang2009preparation}. Together with high-resolution single-site\cite{bakr2010probing,sherson2010single} and single-spin manipulation\cite{weitenberg2011single}, it may lead to one-way quantum computation\cite{walther2005experimental,raussendorf2001one}. Owing to the site-resolved spin addressability in the spin-dependent superlattice, abundant physics of many-body systems can be studied with the present setup, e.g. ring exchange in a spin plaquette\cite{paredes2008minimum} of the minimum instance for the Kitaev model\cite{kitaev2003fault} and quantum magnetism of spin systems\cite{stamper2013spinor}.

During preparation of the manuscript, we became aware of a recent related work by T. Fukuhara et al.\cite{fukuhara2015spatially} detecting a spin-entanglement wave in a Bose-Hubbard chain.

% If in two-column mode, this environment will change to single-column
% format so that long equations can be displayed. Use
% sparingly.
%\begin{widetext}
% put long equation here
%\end{widetext}

% If you have acknowledgments, this puts in the proper section head.
%\begin{acknowledgments}
% put your acknowledgments here.
%\end{acknowledgments}

% Create the reference section using BibTeX:
\bibliography{reference}

\begin{thebibliography}{37}
\expandafter\ifx\csname natexlab\endcsname\relax\def\natexlab#1{#1}\fi
\expandafter\ifx\csname bibnamefont\endcsname\relax
  \def\bibnamefont#1{#1}\fi
\expandafter\ifx\csname bibfnamefont\endcsname\relax
  \def\bibfnamefont#1{#1}\fi
\expandafter\ifx\csname citenamefont\endcsname\relax
  \def\citenamefont#1{#1}\fi
\expandafter\ifx\csname url\endcsname\relax
  \def\url#1{\texttt{#1}}\fi
\expandafter\ifx\csname urlprefix\endcsname\relax\def\urlprefix{URL }\fi
\providecommand{\bibinfo}[2]{#2}
\providecommand{\eprint}[2][]{\url{#2}}

\bibitem[{\citenamefont{Nielsen and Chuang}(2010)}]{nielsen2010quantum}
\bibinfo{author}{\bibfnamefont{M.~A.} \bibnamefont{Nielsen}} \bibnamefont{and}
  \bibinfo{author}{\bibfnamefont{I.~L.} \bibnamefont{Chuang}},
  \emph{\bibinfo{title}{Quantum computation and quantum information}}
  (\bibinfo{publisher}{Cambridge university press}, \bibinfo{year}{2010}).

\bibitem[{\citenamefont{Pan et~al.}(2012)\citenamefont{Pan, Chen, Lu,
  Weinfurter, Zeilinger, and {\.Z}ukowski}}]{pan2012multiphoton}
\bibinfo{author}{\bibfnamefont{J.-W.} \bibnamefont{Pan}},
  \bibinfo{author}{\bibfnamefont{Z.-B.} \bibnamefont{Chen}},
  \bibinfo{author}{\bibfnamefont{C.-Y.} \bibnamefont{Lu}},
  \bibinfo{author}{\bibfnamefont{H.}~\bibnamefont{Weinfurter}},
  \bibinfo{author}{\bibfnamefont{A.}~\bibnamefont{Zeilinger}},
  \bibnamefont{and}
  \bibinfo{author}{\bibfnamefont{M.}~\bibnamefont{{\.Z}ukowski}},
  \bibinfo{journal}{Rev. Mod. Phys.} \textbf{\bibinfo{volume}{84}},
  \bibinfo{pages}{777} (\bibinfo{year}{2012}).

\bibitem[{\citenamefont{Blatt and Wineland}(2008)}]{blatt2008entangled}
\bibinfo{author}{\bibfnamefont{R.}~\bibnamefont{Blatt}} \bibnamefont{and}
  \bibinfo{author}{\bibfnamefont{D.}~\bibnamefont{Wineland}},
  \bibinfo{journal}{Nature} \textbf{\bibinfo{volume}{453}},
  \bibinfo{pages}{1008} (\bibinfo{year}{2008}).

\bibitem[{\citenamefont{Devoret and
  Schoelkopf}(2013)}]{devoret2013superconducting}
\bibinfo{author}{\bibfnamefont{M.}~\bibnamefont{Devoret}} \bibnamefont{and}
  \bibinfo{author}{\bibfnamefont{R.}~\bibnamefont{Schoelkopf}},
  \bibinfo{journal}{Science} \textbf{\bibinfo{volume}{339}},
  \bibinfo{pages}{1169} (\bibinfo{year}{2013}).

\bibitem[{\citenamefont{Awschalom et~al.}(2013)\citenamefont{Awschalom,
  Bassett, Dzurak, Hu, and Petta}}]{awschalom2013quantum}
\bibinfo{author}{\bibfnamefont{D.~D.} \bibnamefont{Awschalom}},
  \bibinfo{author}{\bibfnamefont{L.~C.} \bibnamefont{Bassett}},
  \bibinfo{author}{\bibfnamefont{A.~S.} \bibnamefont{Dzurak}},
  \bibinfo{author}{\bibfnamefont{E.~L.} \bibnamefont{Hu}}, \bibnamefont{and}
  \bibinfo{author}{\bibfnamefont{J.~R.} \bibnamefont{Petta}},
  \bibinfo{journal}{Science} \textbf{\bibinfo{volume}{339}},
  \bibinfo{pages}{1174} (\bibinfo{year}{2013}).

\bibitem[{\citenamefont{Ladd et~al.}(2010)\citenamefont{Ladd, Jelezko,
  Laflamme, Nakamura, Monroe, and O’Brien}}]{ladd2010quantum}
\bibinfo{author}{\bibfnamefont{T.~D.} \bibnamefont{Ladd}},
  \bibinfo{author}{\bibfnamefont{F.}~\bibnamefont{Jelezko}},
  \bibinfo{author}{\bibfnamefont{R.}~\bibnamefont{Laflamme}},
  \bibinfo{author}{\bibfnamefont{Y.}~\bibnamefont{Nakamura}},
  \bibinfo{author}{\bibfnamefont{C.}~\bibnamefont{Monroe}}, \bibnamefont{and}
  \bibinfo{author}{\bibfnamefont{J.~L.} \bibnamefont{O’Brien}},
  \bibinfo{journal}{Nature} \textbf{\bibinfo{volume}{464}}, \bibinfo{pages}{45}
  (\bibinfo{year}{2010}).

\bibitem[{\citenamefont{Trabesinger}(2012)}]{trabesinger2012quantum}
\bibinfo{author}{\bibfnamefont{A.}~\bibnamefont{Trabesinger}},
  \bibinfo{journal}{Nat. Phys.} \textbf{\bibinfo{volume}{8}},
  \bibinfo{pages}{263} (\bibinfo{year}{2012}).

\bibitem[{\citenamefont{Bloch}(2008)}]{bloch2008quantum}
\bibinfo{author}{\bibfnamefont{I.}~\bibnamefont{Bloch}},
  \bibinfo{journal}{Nature} \textbf{\bibinfo{volume}{453}},
  \bibinfo{pages}{1016} (\bibinfo{year}{2008}).

\bibitem[{\citenamefont{Vaucher et~al.}(2008)\citenamefont{Vaucher, Nunnenkamp,
  and Jaksch}}]{vaucher2008creation}
\bibinfo{author}{\bibfnamefont{B.}~\bibnamefont{Vaucher}},
  \bibinfo{author}{\bibfnamefont{A.}~\bibnamefont{Nunnenkamp}},
  \bibnamefont{and} \bibinfo{author}{\bibfnamefont{D.}~\bibnamefont{Jaksch}},
  \bibinfo{journal}{New J. Phys.} \textbf{\bibinfo{volume}{10}},
  \bibinfo{pages}{023005} (\bibinfo{year}{2008}).

\bibitem[{\citenamefont{Raussendorf and Briegel}(2001)}]{raussendorf2001one}
\bibinfo{author}{\bibfnamefont{R.}~\bibnamefont{Raussendorf}} \bibnamefont{and}
  \bibinfo{author}{\bibfnamefont{H.~J.} \bibnamefont{Briegel}},
  \bibinfo{journal}{Phys. Rev. Lett.} \textbf{\bibinfo{volume}{86}},
  \bibinfo{pages}{5188} (\bibinfo{year}{2001}).

\bibitem[{\citenamefont{Walther et~al.}(2005)\citenamefont{Walther, Resch,
  Rudolph, Schenck, Weinfurter, Vedral, Aspelmeyer, and
  Zeilinger}}]{walther2005experimental}
\bibinfo{author}{\bibfnamefont{P.}~\bibnamefont{Walther}},
  \bibinfo{author}{\bibfnamefont{K.~J.} \bibnamefont{Resch}},
  \bibinfo{author}{\bibfnamefont{T.}~\bibnamefont{Rudolph}},
  \bibinfo{author}{\bibfnamefont{E.}~\bibnamefont{Schenck}},
  \bibinfo{author}{\bibfnamefont{H.}~\bibnamefont{Weinfurter}},
  \bibinfo{author}{\bibfnamefont{V.}~\bibnamefont{Vedral}},
  \bibinfo{author}{\bibfnamefont{M.}~\bibnamefont{Aspelmeyer}},
  \bibnamefont{and}
  \bibinfo{author}{\bibfnamefont{A.}~\bibnamefont{Zeilinger}},
  \bibinfo{journal}{Nature} \textbf{\bibinfo{volume}{434}},
  \bibinfo{pages}{169} (\bibinfo{year}{2005}).

\bibitem[{\citenamefont{Duan et~al.}(2003)\citenamefont{Duan, Demler, and
  Lukin}}]{duan2003controlling}
\bibinfo{author}{\bibfnamefont{L.-M.} \bibnamefont{Duan}},
  \bibinfo{author}{\bibfnamefont{E.}~\bibnamefont{Demler}}, \bibnamefont{and}
  \bibinfo{author}{\bibfnamefont{M.}~\bibnamefont{Lukin}},
  \bibinfo{journal}{Phys. Rev. Lett.} \textbf{\bibinfo{volume}{91}},
  \bibinfo{pages}{090402} (\bibinfo{year}{2003}).

\bibitem[{\citenamefont{Trotzky et~al.}(2008)\citenamefont{Trotzky, Cheinet,
  F{\"o}lling, Feld, Schnorrberger, Rey, Polkovnikov, Demler, Lukin, and
  Bloch}}]{trotzky2008time}
\bibinfo{author}{\bibfnamefont{S.}~\bibnamefont{Trotzky}},
  \bibinfo{author}{\bibfnamefont{P.}~\bibnamefont{Cheinet}},
  \bibinfo{author}{\bibfnamefont{S.}~\bibnamefont{F{\"o}lling}},
  \bibinfo{author}{\bibfnamefont{M.}~\bibnamefont{Feld}},
  \bibinfo{author}{\bibfnamefont{U.}~\bibnamefont{Schnorrberger}},
  \bibinfo{author}{\bibfnamefont{A.~M.} \bibnamefont{Rey}},
  \bibinfo{author}{\bibfnamefont{A.}~\bibnamefont{Polkovnikov}},
  \bibinfo{author}{\bibfnamefont{E.}~\bibnamefont{Demler}},
  \bibinfo{author}{\bibfnamefont{M.}~\bibnamefont{Lukin}}, \bibnamefont{and}
  \bibinfo{author}{\bibfnamefont{I.}~\bibnamefont{Bloch}},
  \bibinfo{journal}{Science} \textbf{\bibinfo{volume}{319}},
  \bibinfo{pages}{295} (\bibinfo{year}{2008}).

\bibitem[{\citenamefont{Anderlini et~al.}(2007)\citenamefont{Anderlini, Lee,
  Brown, Sebby-Strabley, Phillips, and Porto}}]{anderlini2007controlled}
\bibinfo{author}{\bibfnamefont{M.}~\bibnamefont{Anderlini}},
  \bibinfo{author}{\bibfnamefont{P.~J.} \bibnamefont{Lee}},
  \bibinfo{author}{\bibfnamefont{B.~L.} \bibnamefont{Brown}},
  \bibinfo{author}{\bibfnamefont{J.}~\bibnamefont{Sebby-Strabley}},
  \bibinfo{author}{\bibfnamefont{W.~D.} \bibnamefont{Phillips}},
  \bibnamefont{and} \bibinfo{author}{\bibfnamefont{J.}~\bibnamefont{Porto}},
  \bibinfo{journal}{Nature} \textbf{\bibinfo{volume}{448}},
  \bibinfo{pages}{452} (\bibinfo{year}{2007}).

\bibitem[{\citenamefont{Julienne et~al.}(1997)\citenamefont{Julienne, Mies,
  Tiesinga, and Williams}}]{julienne1997collisional}
\bibinfo{author}{\bibfnamefont{P.}~\bibnamefont{Julienne}},
  \bibinfo{author}{\bibfnamefont{F.}~\bibnamefont{Mies}},
  \bibinfo{author}{\bibfnamefont{E.}~\bibnamefont{Tiesinga}}, \bibnamefont{and}
  \bibinfo{author}{\bibfnamefont{C.}~\bibnamefont{Williams}},
  \bibinfo{journal}{Phys. Rev. Lett.} \textbf{\bibinfo{volume}{78}},
  \bibinfo{pages}{1880} (\bibinfo{year}{1997}).

\bibitem[{\citenamefont{Schmaljohann et~al.}(2004)\citenamefont{Schmaljohann,
  Erhard, Kronj\"ager, Kottke, Van~Staa, Cacciapuoti, Arlt, Bongs, and
  Sengstock}}]{schmaljohann2004dynamics}
\bibinfo{author}{\bibfnamefont{H.}~\bibnamefont{Schmaljohann}},
  \bibinfo{author}{\bibfnamefont{M.}~\bibnamefont{Erhard}},
  \bibinfo{author}{\bibfnamefont{J.}~\bibnamefont{Kronj\"ager}},
  \bibinfo{author}{\bibfnamefont{M.}~\bibnamefont{Kottke}},
  \bibinfo{author}{\bibfnamefont{S.}~\bibnamefont{Van~Staa}},
  \bibinfo{author}{\bibfnamefont{L.}~\bibnamefont{Cacciapuoti}},
  \bibinfo{author}{\bibfnamefont{J.}~\bibnamefont{Arlt}},
  \bibinfo{author}{\bibfnamefont{K.}~\bibnamefont{Bongs}}, \bibnamefont{and}
  \bibinfo{author}{\bibfnamefont{K.}~\bibnamefont{Sengstock}},
  \bibinfo{journal}{Phys. Rev. Lett.} \textbf{\bibinfo{volume}{92}},
  \bibinfo{pages}{040402} (\bibinfo{year}{2004}).

\bibitem[{\citenamefont{Clauser et~al.}(1969)\citenamefont{Clauser, Horne,
  Shimony, and Holt}}]{clauser1969proposed}
\bibinfo{author}{\bibfnamefont{J.~F.} \bibnamefont{Clauser}},
  \bibinfo{author}{\bibfnamefont{M.~A.} \bibnamefont{Horne}},
  \bibinfo{author}{\bibfnamefont{A.}~\bibnamefont{Shimony}}, \bibnamefont{and}
  \bibinfo{author}{\bibfnamefont{R.~A.} \bibnamefont{Holt}},
  \bibinfo{journal}{Phys. Rev. Lett.} \textbf{\bibinfo{volume}{23}},
  \bibinfo{pages}{880} (\bibinfo{year}{1969}).

\bibitem[{\citenamefont{Bloch et~al.}(2008)\citenamefont{Bloch, Dalibard, and
  Zwerger}}]{bloch2008many}
\bibinfo{author}{\bibfnamefont{I.}~\bibnamefont{Bloch}},
  \bibinfo{author}{\bibfnamefont{J.}~\bibnamefont{Dalibard}}, \bibnamefont{and}
  \bibinfo{author}{\bibfnamefont{W.}~\bibnamefont{Zwerger}},
  \bibinfo{journal}{Rev. Mod. Phys.} \textbf{\bibinfo{volume}{80}},
  \bibinfo{pages}{885} (\bibinfo{year}{2008}).

\bibitem[{\citenamefont{Lewenstein et~al.}(2012)\citenamefont{Lewenstein,
  Sanpera, and Ahufinger}}]{lewenstein2012ultracold}
\bibinfo{author}{\bibfnamefont{M.}~\bibnamefont{Lewenstein}},
  \bibinfo{author}{\bibfnamefont{A.}~\bibnamefont{Sanpera}}, \bibnamefont{and}
  \bibinfo{author}{\bibfnamefont{V.}~\bibnamefont{Ahufinger}},
  \emph{\bibinfo{title}{Ultracold Atoms in Optical Lattices: Simulating quantum
  many-body systems}} (\bibinfo{publisher}{OUP Oxford}, \bibinfo{year}{2012}).

\bibitem[{\citenamefont{Dalfovo et~al.}(1999)\citenamefont{Dalfovo, Giorgini,
  Pitaevskii, and Stringari}}]{dalfovo1999theory}
\bibinfo{author}{\bibfnamefont{F.}~\bibnamefont{Dalfovo}},
  \bibinfo{author}{\bibfnamefont{S.}~\bibnamefont{Giorgini}},
  \bibinfo{author}{\bibfnamefont{L.~P.} \bibnamefont{Pitaevskii}},
  \bibnamefont{and}
  \bibinfo{author}{\bibfnamefont{S.}~\bibnamefont{Stringari}},
  \bibinfo{journal}{Rev. Mod. Phys.} \textbf{\bibinfo{volume}{71}},
  \bibinfo{pages}{463} (\bibinfo{year}{1999}).

\bibitem[{\citenamefont{Jaksch et~al.}(1998)\citenamefont{Jaksch, Bruder,
  Cirac, Gardiner, and Zoller}}]{jaksch1998cold}
\bibinfo{author}{\bibfnamefont{D.}~\bibnamefont{Jaksch}},
  \bibinfo{author}{\bibfnamefont{C.}~\bibnamefont{Bruder}},
  \bibinfo{author}{\bibfnamefont{J.~I.} \bibnamefont{Cirac}},
  \bibinfo{author}{\bibfnamefont{C.~W.} \bibnamefont{Gardiner}},
  \bibnamefont{and} \bibinfo{author}{\bibfnamefont{P.}~\bibnamefont{Zoller}},
  \bibinfo{journal}{Phys. Rev. Lett.} \textbf{\bibinfo{volume}{81}},
  \bibinfo{pages}{3108} (\bibinfo{year}{1998}).

\bibitem[{\citenamefont{Greiner et~al.}(2002)\citenamefont{Greiner, Mandel,
  Esslinger, H\"ansch, and Bloch}}]{greiner2002quantum}
\bibinfo{author}{\bibfnamefont{M.}~\bibnamefont{Greiner}},
  \bibinfo{author}{\bibfnamefont{O.}~\bibnamefont{Mandel}},
  \bibinfo{author}{\bibfnamefont{T.}~\bibnamefont{Esslinger}},
  \bibinfo{author}{\bibfnamefont{T.~W.} \bibnamefont{H\"ansch}},
  \bibnamefont{and} \bibinfo{author}{\bibfnamefont{I.}~\bibnamefont{Bloch}},
  \bibinfo{journal}{Nature} \textbf{\bibinfo{volume}{415}}, \bibinfo{pages}{39}
  (\bibinfo{year}{2002}).

\bibitem[{\citenamefont{J{\"o}rdens et~al.}(2008)\citenamefont{J{\"o}rdens,
  Strohmaier, G{\"u}nter, Moritz, and Esslinger}}]{jordens2008mott}
\bibinfo{author}{\bibfnamefont{R.}~\bibnamefont{J{\"o}rdens}},
  \bibinfo{author}{\bibfnamefont{N.}~\bibnamefont{Strohmaier}},
  \bibinfo{author}{\bibfnamefont{K.}~\bibnamefont{G{\"u}nter}},
  \bibinfo{author}{\bibfnamefont{H.}~\bibnamefont{Moritz}}, \bibnamefont{and}
  \bibinfo{author}{\bibfnamefont{T.}~\bibnamefont{Esslinger}},
  \bibinfo{journal}{Nature} \textbf{\bibinfo{volume}{455}},
  \bibinfo{pages}{204} (\bibinfo{year}{2008}).

\bibitem[{\citenamefont{Schneider et~al.}(2008)\citenamefont{Schneider,
  Hackerm{\"u}ller, Will, Best, Bloch, Costi, Helmes, Rasch, and
  Rosch}}]{schneider2008metallic}
\bibinfo{author}{\bibfnamefont{U.}~\bibnamefont{Schneider}},
  \bibinfo{author}{\bibfnamefont{L.}~\bibnamefont{Hackerm{\"u}ller}},
  \bibinfo{author}{\bibfnamefont{S.}~\bibnamefont{Will}},
  \bibinfo{author}{\bibfnamefont{T.}~\bibnamefont{Best}},
  \bibinfo{author}{\bibfnamefont{I.}~\bibnamefont{Bloch}},
  \bibinfo{author}{\bibfnamefont{T.}~\bibnamefont{Costi}},
  \bibinfo{author}{\bibfnamefont{R.}~\bibnamefont{Helmes}},
  \bibinfo{author}{\bibfnamefont{D.}~\bibnamefont{Rasch}}, \bibnamefont{and}
  \bibinfo{author}{\bibfnamefont{A.}~\bibnamefont{Rosch}},
  \bibinfo{journal}{Science} \textbf{\bibinfo{volume}{322}},
  \bibinfo{pages}{1520} (\bibinfo{year}{2008}).

\bibitem[{Ent()}]{EntangleSuperlatticeSI}
\emph{\bibinfo{title}{Supplementary material}}.

\bibitem[{\citenamefont{Trotzky et~al.}(2010)\citenamefont{Trotzky, Chen,
  Schnorrberger, Cheinet, and Bloch}}]{trotzky2010controlling}
\bibinfo{author}{\bibfnamefont{S.}~\bibnamefont{Trotzky}},
  \bibinfo{author}{\bibfnamefont{Y.-A.} \bibnamefont{Chen}},
  \bibinfo{author}{\bibfnamefont{U.}~\bibnamefont{Schnorrberger}},
  \bibinfo{author}{\bibfnamefont{P.}~\bibnamefont{Cheinet}}, \bibnamefont{and}
  \bibinfo{author}{\bibfnamefont{I.}~\bibnamefont{Bloch}},
  \bibinfo{journal}{Phys. Rev. Lett.} \textbf{\bibinfo{volume}{105}},
  \bibinfo{pages}{265303} (\bibinfo{year}{2010}).

\bibitem[{\citenamefont{Blinov et~al.}(2004)\citenamefont{Blinov, Moehring,
  Duan, and Monroe}}]{blinov2004observation}
\bibinfo{author}{\bibfnamefont{B.}~\bibnamefont{Blinov}},
  \bibinfo{author}{\bibfnamefont{D.}~\bibnamefont{Moehring}},
  \bibinfo{author}{\bibfnamefont{L.-M.} \bibnamefont{Duan}}, \bibnamefont{and}
  \bibinfo{author}{\bibfnamefont{C.}~\bibnamefont{Monroe}},
  \bibinfo{journal}{Nature} \textbf{\bibinfo{volume}{428}},
  \bibinfo{pages}{153} (\bibinfo{year}{2004}).

\bibitem[{\citenamefont{Bakr et~al.}(2011)\citenamefont{Bakr, Preiss, Tai, Ma,
  Simon, and Greiner}}]{bakr2011orbital}
\bibinfo{author}{\bibfnamefont{W.~S.} \bibnamefont{Bakr}},
  \bibinfo{author}{\bibfnamefont{P.~M.} \bibnamefont{Preiss}},
  \bibinfo{author}{\bibfnamefont{M.~E.} \bibnamefont{Tai}},
  \bibinfo{author}{\bibfnamefont{R.}~\bibnamefont{Ma}},
  \bibinfo{author}{\bibfnamefont{J.}~\bibnamefont{Simon}}, \bibnamefont{and}
  \bibinfo{author}{\bibfnamefont{M.}~\bibnamefont{Greiner}},
  \bibinfo{journal}{Nature} \textbf{\bibinfo{volume}{480}},
  \bibinfo{pages}{500} (\bibinfo{year}{2011}).

\bibitem[{\citenamefont{Alves and Jaksch}(2004)}]{alves2004multipartite}
\bibinfo{author}{\bibfnamefont{C.~M.} \bibnamefont{Alves}} \bibnamefont{and}
  \bibinfo{author}{\bibfnamefont{D.}~\bibnamefont{Jaksch}},
  \bibinfo{journal}{Phys. Rev. Lett.} \textbf{\bibinfo{volume}{93}},
  \bibinfo{pages}{110501} (\bibinfo{year}{2004}).

\bibitem[{\citenamefont{Jiang et~al.}(2009)\citenamefont{Jiang, Rey,
  Romero-Isart, Garc{\'\i}a-Ripoll, Sanpera, and Lukin}}]{jiang2009preparation}
\bibinfo{author}{\bibfnamefont{L.}~\bibnamefont{Jiang}},
  \bibinfo{author}{\bibfnamefont{A.~M.} \bibnamefont{Rey}},
  \bibinfo{author}{\bibfnamefont{O.}~\bibnamefont{Romero-Isart}},
  \bibinfo{author}{\bibfnamefont{J.~J.} \bibnamefont{Garc{\'\i}a-Ripoll}},
  \bibinfo{author}{\bibfnamefont{A.}~\bibnamefont{Sanpera}}, \bibnamefont{and}
  \bibinfo{author}{\bibfnamefont{M.~D.} \bibnamefont{Lukin}},
  \bibinfo{journal}{Phys. Rev. A} \textbf{\bibinfo{volume}{79}},
  \bibinfo{pages}{022309} (\bibinfo{year}{2009}).

\bibitem[{\citenamefont{Bakr et~al.}(2010)\citenamefont{Bakr, Peng, Tai, Ma,
  Simon, Gillen, Foelling, Pollet, and Greiner}}]{bakr2010probing}
\bibinfo{author}{\bibfnamefont{W.~S.} \bibnamefont{Bakr}},
  \bibinfo{author}{\bibfnamefont{A.}~\bibnamefont{Peng}},
  \bibinfo{author}{\bibfnamefont{M.~E.} \bibnamefont{Tai}},
  \bibinfo{author}{\bibfnamefont{R.}~\bibnamefont{Ma}},
  \bibinfo{author}{\bibfnamefont{J.}~\bibnamefont{Simon}},
  \bibinfo{author}{\bibfnamefont{J.~I.} \bibnamefont{Gillen}},
  \bibinfo{author}{\bibfnamefont{S.}~\bibnamefont{Foelling}},
  \bibinfo{author}{\bibfnamefont{L.}~\bibnamefont{Pollet}}, \bibnamefont{and}
  \bibinfo{author}{\bibfnamefont{M.}~\bibnamefont{Greiner}},
  \bibinfo{journal}{Science} \textbf{\bibinfo{volume}{329}},
  \bibinfo{pages}{547} (\bibinfo{year}{2010}).

\bibitem[{\citenamefont{Sherson et~al.}(2010)\citenamefont{Sherson, Weitenberg,
  Endres, Cheneau, Bloch, and Kuhr}}]{sherson2010single}
\bibinfo{author}{\bibfnamefont{J.~F.} \bibnamefont{Sherson}},
  \bibinfo{author}{\bibfnamefont{C.}~\bibnamefont{Weitenberg}},
  \bibinfo{author}{\bibfnamefont{M.}~\bibnamefont{Endres}},
  \bibinfo{author}{\bibfnamefont{M.}~\bibnamefont{Cheneau}},
  \bibinfo{author}{\bibfnamefont{I.}~\bibnamefont{Bloch}}, \bibnamefont{and}
  \bibinfo{author}{\bibfnamefont{S.}~\bibnamefont{Kuhr}},
  \bibinfo{journal}{Nature} \textbf{\bibinfo{volume}{467}}, \bibinfo{pages}{68}
  (\bibinfo{year}{2010}).

\bibitem[{\citenamefont{Weitenberg et~al.}(2011)\citenamefont{Weitenberg,
  Endres, Sherson, Cheneau, Schauss, Fukuhara, Bloch, and
  Kuhr}}]{weitenberg2011single}
\bibinfo{author}{\bibfnamefont{C.}~\bibnamefont{Weitenberg}},
  \bibinfo{author}{\bibfnamefont{M.}~\bibnamefont{Endres}},
  \bibinfo{author}{\bibfnamefont{J.~F.} \bibnamefont{Sherson}},
  \bibinfo{author}{\bibfnamefont{M.}~\bibnamefont{Cheneau}},
  \bibinfo{author}{\bibfnamefont{P.}~\bibnamefont{Schauss}},
  \bibinfo{author}{\bibfnamefont{T.}~\bibnamefont{Fukuhara}},
  \bibinfo{author}{\bibfnamefont{I.}~\bibnamefont{Bloch}}, \bibnamefont{and}
  \bibinfo{author}{\bibfnamefont{S.}~\bibnamefont{Kuhr}},
  \bibinfo{journal}{Nature} \textbf{\bibinfo{volume}{471}},
  \bibinfo{pages}{319} (\bibinfo{year}{2011}).

\bibitem[{\citenamefont{Paredes and Bloch}(2008)}]{paredes2008minimum}
\bibinfo{author}{\bibfnamefont{B.}~\bibnamefont{Paredes}} \bibnamefont{and}
  \bibinfo{author}{\bibfnamefont{I.}~\bibnamefont{Bloch}},
  \bibinfo{journal}{Phys. Rev. A} \textbf{\bibinfo{volume}{77}},
  \bibinfo{pages}{023603} (\bibinfo{year}{2008}).

\bibitem[{\citenamefont{Kitaev}(2003)}]{kitaev2003fault}
\bibinfo{author}{\bibfnamefont{A.~Y.} \bibnamefont{Kitaev}},
  \bibinfo{journal}{Ann. Phys.} \textbf{\bibinfo{volume}{303}},
  \bibinfo{pages}{2} (\bibinfo{year}{2003}).

\bibitem[{\citenamefont{Stamper-Kurn and Ueda}(2013)}]{stamper2013spinor}
\bibinfo{author}{\bibfnamefont{D.~M.} \bibnamefont{Stamper-Kurn}}
  \bibnamefont{and} \bibinfo{author}{\bibfnamefont{M.}~\bibnamefont{Ueda}},
  \bibinfo{journal}{Rev. Mod. Phys.} \textbf{\bibinfo{volume}{85}},
  \bibinfo{pages}{1191} (\bibinfo{year}{2013}).

\bibitem[{\citenamefont{Fukuhara et~al.}(2015)\citenamefont{Fukuhara, Hild,
  Zeiher, Schau\ss{}, Bloch, Endres, and Gross}}]{fukuhara2015spatially}
\bibinfo{author}{\bibfnamefont{T.}~\bibnamefont{Fukuhara}},
  \bibinfo{author}{\bibfnamefont{S.}~\bibnamefont{Hild}},
  \bibinfo{author}{\bibfnamefont{J.}~\bibnamefont{Zeiher}},
  \bibinfo{author}{\bibfnamefont{P.}~\bibnamefont{Schau\ss{}}},
  \bibinfo{author}{\bibfnamefont{I.}~\bibnamefont{Bloch}},
  \bibinfo{author}{\bibfnamefont{M.}~\bibnamefont{Endres}}, \bibnamefont{and}
  \bibinfo{author}{\bibfnamefont{C.}~\bibnamefont{Gross}},
  \bibinfo{journal}{Phys. Rev. Lett.} \textbf{\bibinfo{volume}{115}},
  \bibinfo{pages}{035302} (\bibinfo{year}{2015}).

\end{thebibliography}
%%%%%%%%%% Merge with supplemental materials %%%%%%%%%%

\widetext
\clearpage

\section*{\large Supplemental Materials}

%%%%%%%%%% Merge with supplemental materials %%%%%%%%%%
%%%%%%%%%% Prefix a "S" to all equations, figures, tables and reset the counter %%%%%%%%%%
\setcounter{equation}{0}
\setcounter{figure}{0}
\setcounter{table}{0}
\setcounter{page}{1}
\makeatletter
\renewcommand{\theequation}{S\arabic{equation}}
\renewcommand{\thefigure}{S\arabic{figure}}
\renewcommand{\bibnumfmt}[1]{[S#1]}
\renewcommand{\citenumfont}[1]{S#1}
%%%%%%%%%% Prefix a "S" to all equations, figures, tables and reset the counter %%%%%%%%%%

\subsection*{Preparing balanced DWs}
The superlattice potential along the X direction is \hbox{$V_x(x)\!=\!V_{xs}\cos^2(k_x x)\!-\!V_{xl}\cos^2(k_x x/2 +\varphi)$}, when the EOM modulation is off. By beating the up-converted light of the 1534-nm laser with the 767-nm laser, the relative frequency of the two lasers are locked at about 5.5 GHz with a linewidth less than 400 kHz in the frequency offset locking scheme. Therefore the relative phase $\varphi$ between the long lattice and the short lattice can be continuously adjusted from 0 to $\pi$ by tuning the offset locking frequency. During the measurement of the superexchange, STO and entanglement, the superlattice phase is tuned to 0.

\subsection*{Spin-dependent superlattices and site-resolved spin manipulation}
The spin-dependent superlattice is realized by overlapping a spin-dependent short lattice with a normal long lattice, as shown in the setup, Fig.1(a) in the matin text. The spin dependent short lattice is created by modulating the polarizations of the forward and the retro-reflected beam. The initial polarization of the forward beam is linear polarized in the horizontal plane (X-Z plane), and it is modulated by the control voltage applied to the electro-optical modulator (EOM). The axis of the EOM is aligned to the horizontal plane with an angle of 45$^{\circ}$ and a quarter-wave plate (QWP) with its fast axis in the horizontal plane is used to adjust the circular polarization induced by the EOM. Then the beam after the QWP is still linear polarized, but rotated by an angle of $\theta/2$ with respect to the original polarization, where $\theta$ is the modulated phase by the EOM voltage. The two forward beams of  short lattice and long lattice (horizontal polarized) are overlapped on a dichroic mirror and focused to the atoms by an achromatic lens. For the retro-reflected beams, a special designed QWP is used, which acts as a QWP for the short lattice while inducing no phase change for the long lattice. The fast axis of this QWP is also aligned in the horizontal plane. The retro-reflected beam of the short lattice is linear polarized with an angle of $-\theta/2$ to the horizontal plane, and $-\theta$ to the forward beam  (Fig.S1(a)), while the retro beam of the long lattice is still horizontal polarized. Thus the interference of the short lattice can be modulated, and the atoms in the spin down (\hbox{$|\!\downarrow\rangle=5S_{1/2}|F=1,m_F=-1\rangle$}) and spin up (\hbox{$|\!\uparrow\rangle=5S_{1/2}|F=2,m_F=-2\rangle$}) state will respond differently due to the different dipole couplings to the left/right circular polarization components of the lattice. The spin-dependent term is proportional to \hbox{$i(\textbf{E}^{\ast}\!\times\!\textbf{E})\!\cdot\!\bm{\mu}$}, where $\textbf{E}$ is the direction of electric filed and $\bm{\mu}$ is the magnetic momentum[S1]. Therefore the largest spin-dependent effect is present when the quantum axis is aligned along the X direction. The modulated lattices along X direction for the two spin states can be described as

\begin{figure}[!b]
\centerline{\includegraphics[width=12.21 cm]{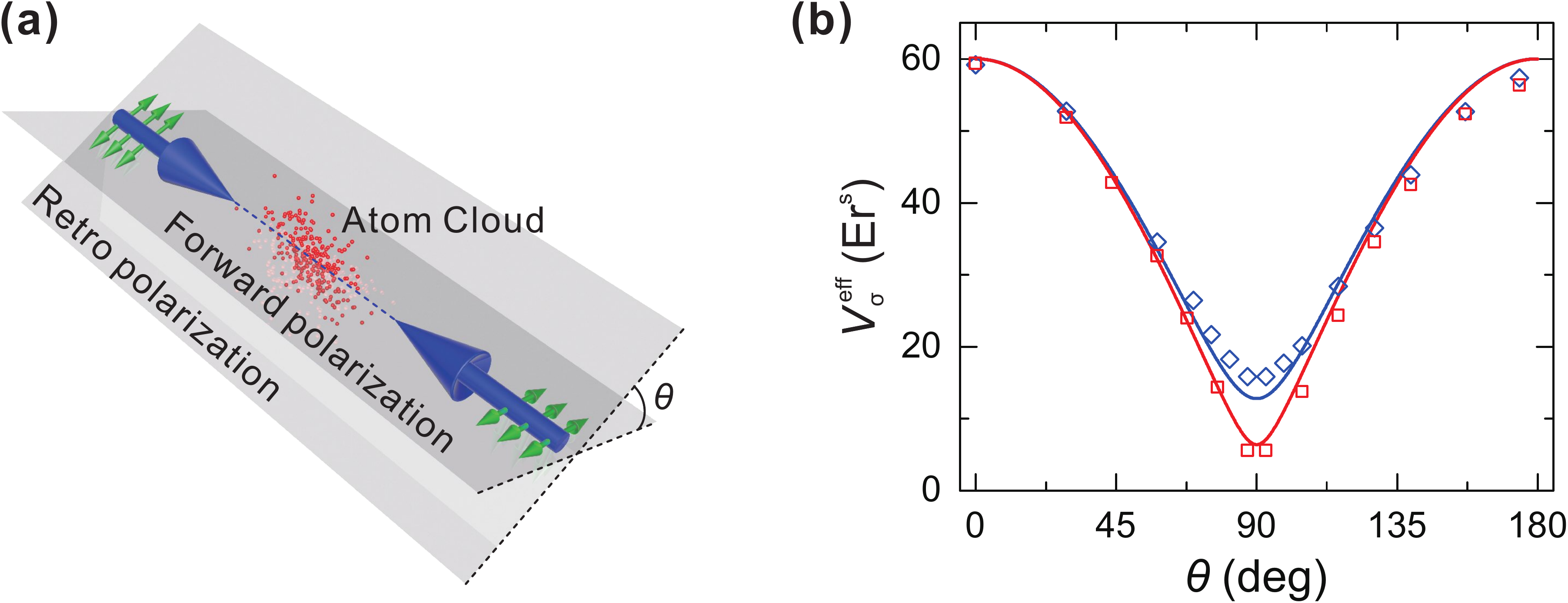}}%
\caption{{\bf Spin dependent short lattice.} \textbf{(a)} Schematics of the linear polarized forward and retro reflected beams with polarization difference of $\theta$, which is modulated by an EOM and QWPs; \textbf{(b)} Effective short lattice depth varied by the EOM modulation for a 60 Er short lattice. The calibration is done by applying a short time parametric modulation to the lattice intensity when the long lattice is off. This confirms the theoretical predications (solid lines) of $V^{\mathrm{eff}}_{\sigma}(\theta) = V_{xs,\sigma} \sqrt{\cos^2\theta+\mathcal{B}_{\sigma}\sin^2\theta}$, $\sigma=|\!\uparrow\rangle$ (blue) or $|\!\downarrow\rangle$ (red).}
\end{figure}

\begin{equation}
V_{\sigma}(x,\theta)  = V_{xs,\sigma}\left[ \mathcal{A}^{+}_{\sigma} \cos^2(k_x x+\frac{\theta}{2})+  \mathcal{A}^{-}_{\sigma} \cos^2(k_x x-\frac{\theta}{2})\right] - V_{xl,\sigma}\cos^2(\frac{k_x}{2} x+\varphi),
\end{equation}
where $\sigma=|\!\uparrow\rangle$ or $|\!\downarrow\rangle$ and $A^{\nu}_\sigma$ is the response for spin $\sigma$ to the circular polarization component of $\nu$ ($+/-$ for right/left circular polarization). Here we have $A^{+}_{\downarrow} = 0.55$,  $A^{-}_{\downarrow} = 0.45$, $A^{+}_{\uparrow} = 0.40$, and  $A^{-}_{\uparrow} = 0.60$, respectively. The effective short lattice  depth will be modulated by the EOM, and it is calibrated by the resonance frequency of the coupling from the ground band to the second excited band. The effective lattice depth for $|\!\downarrow\rangle$ and $|\!\uparrow\rangle$ are depending on the EOM modulating:  \hbox{$V^{\mathrm{eff}}_{\sigma}(\theta) = V_{xs,\sigma} \sqrt{\cos^2\theta+\mathcal{B}_{\sigma}\sin^2\theta}$ }, as shown in Fig.S1(b). In our experiment, $\mathcal{B}_\downarrow = 1.13\times10^{-2}$ and $\mathcal{B}_\uparrow = 4.54\times10^{-2}$.

We find that the superlattice phase will also be modulated by the EOM, since $V_{\sigma}(x,\theta)$ can be written as
\begin{equation}
\begin{split}
V_{\sigma}(x,\theta) =  &V_{xs,\sigma} \cos\theta\ \cos^2(k_x x)  + \frac{V_{xs,\sigma}}{2}\left[1-\cos\theta+(\mathcal{A}^{+}_{\sigma}-\mathcal{A}^{-}_{\sigma})\sin\theta\right]\\
&- V_{xl,\sigma}\cos^2(\frac{k_x}{2} x+\varphi) -   V_{xs,\sigma}\ (\mathcal{A}^{+}_{\sigma}-\mathcal{A}^{-}_{\sigma})\sin\theta \cos^2 (k_x x-\frac{\pi}{4})
\end{split}
\end{equation}
The last term in Eq.2 is the part which will introduce the spin-dependence, and it will also change the superlattice phase $\varphi$. For simplicity, we consider the balanced DWs $\varphi=0$ and an EOM modulation of $0<\theta<90^{\circ}$. Since $\mathcal{A}^{+}_{\downarrow}>\mathcal{A}^{-}_{\downarrow}$, the DWs of $|\!\downarrow\rangle$  will be tilted to the right side, while $\mathcal{A}^{+}_{\uparrow} < \mathcal{A}^{-}_{\uparrow}$,  the DWs of $|\!\uparrow\rangle$  will be tilted to the left side, as shown in Fig.S2(c,d). In this case, the energy shifts for $|\!\uparrow\rangle/|\!\downarrow\rangle$ in left/right sites of the DWs are different. With this method, we can address the left or right sites in the DWs without affecting the other ones.

\begin{figure}[!t]
\centerline{\includegraphics[width=13.467 cm]{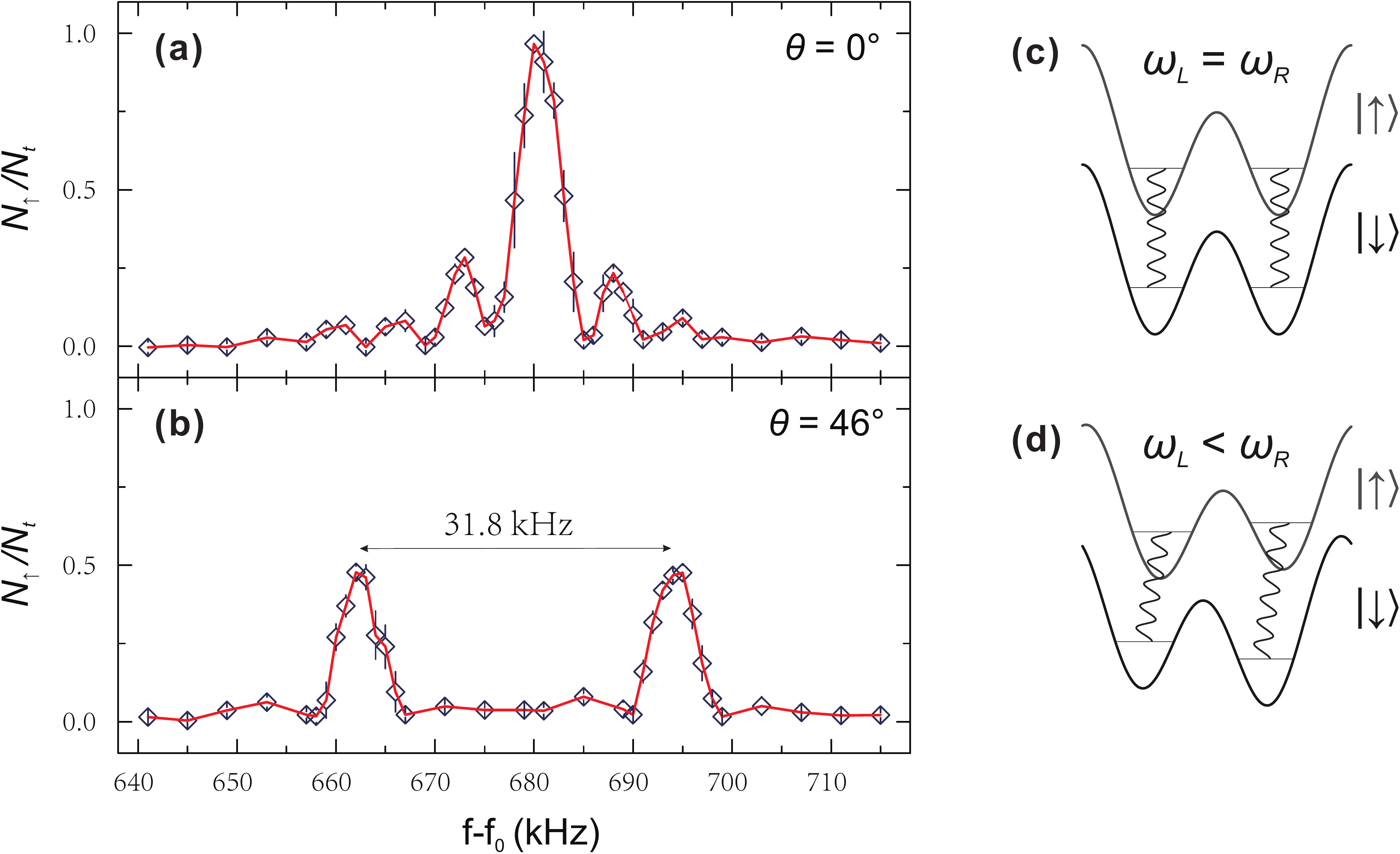}}%
\caption{\noindent {\bf Site-resolved MW addressing spectrum in the spin-dependent superlattice.} Atoms are loaded in lattices $V_{xl}=56.3$ Er, $V_{xs}=150$ Er and $V_{y}=40$ Er  for transition of $|\!\uparrow\rangle\Rightarrow|\!\downarrow\rangle$. \textbf{(a)} When the EOM modulation is off, the spectrum has a normal single maximum peak, which means the left and right sites have the same transition frequency. \textbf{(b)} When the EOM is modulated to an angle of 46$^\circ$, two distinct peaks with 31.8 kHz separation are observed, corresponding to the two microwave frequencies for addressing the left or the right sites. \textbf{(c,d)} The DW potentials are shown when the EOM modulation induces an angle of $\theta=0^\circ$ or $\theta=46^\circ$ between the polarization of the forward and retro-reflected beam in the short lattice.}
\end{figure}
Shown in Fig.S2, two distinct micro-wave spectrums are obtain by a MW pulse with Rabi frequency $\Omega=3$ kHz and pulse length 167 $\mu$s in the case with or without the modulation of EOM. The spectroscopies are measured at lattice depths of $V_{xl}=56.3$ Er, $V_{xs}=150$ Er by recording the number of atoms flipped by the MW. We first measure without EOM modulation, $\theta = 0^{\circ}$, and a single resonance peak can be obtained. While  two separated peaks with separation of 31.8 kHz are observed when we turn the EOM modulation to $\theta = 46^\circ$. The two peaks correspond to the different resonance frequencies for the left and right sites, respectively.

\subsection*{Measure the filling parameters in MI}

After entereing the MI regime, we increase lattice depth on the X and Y direction to freeze the tunneling. Afterwards, we use a rapid adiabatic passage to transfer all the atoms from the initial state $|\!\downarrow\rangle$ to $|\!\uparrow^\prime\rangle$ ($5S_{1/2}|F=2,m_F=-1\rangle$). The atoms in the $F=2$ mainfold can undergo hyperfine changing collisions (Ref.[15] and [16] in the main text), and have a shorter lifetime for the sites filled with two or more atoms. The double fillings will be removed from the trap by this filtering process after a holding time of 500 ms in our experiment. Comparing the atom numbers in the ROI with and without collisional loss, we can obtain the double occupancy as \hbox{$p_2\!\approx\!1.8\%$}, as shown in Fig.S3 (a,b). The average filling in the ROI of the MI is about 0.80, which indicates that multiple occupancies have low probabilities. We will neglect the occupancies with more than two atoms for later calibrations and experiments. After this filtering, we transfer the remaining atoms to the long lattice by ramping down the short lattice and measure the loss after another 500 ms holding time (Fig.S3 (c)). The loss in this process is mainly due to the singly filled states of \hbox{$|\!\uparrow^{\prime},\uparrow^{\prime}\rangle$}. Therefore we measured the occupancy of both DW sites filled with one atom as \hbox{$p_{1|1}\!\approx\!58\%$}. According to this calibration, we know that the filling of $1|1$ in the DWs contain \hbox{$0.58/0.80\!=\!72.5\%$} of the total number of atoms in the ROI. Furthermore, by assuming that the fillings between the neighboring sites are independent (in deep lattices), we derive the filling probability of one atom per site as \hbox{$p_1=\!\sqrt{p_{1|1}} \! = \!76.2\%$}, and therefore the vacancy filling as \hbox{$p_0\! \approx \!22\%$}.
\begin{figure}[!tb]
\centerline{\includegraphics[width=13.143 cm]{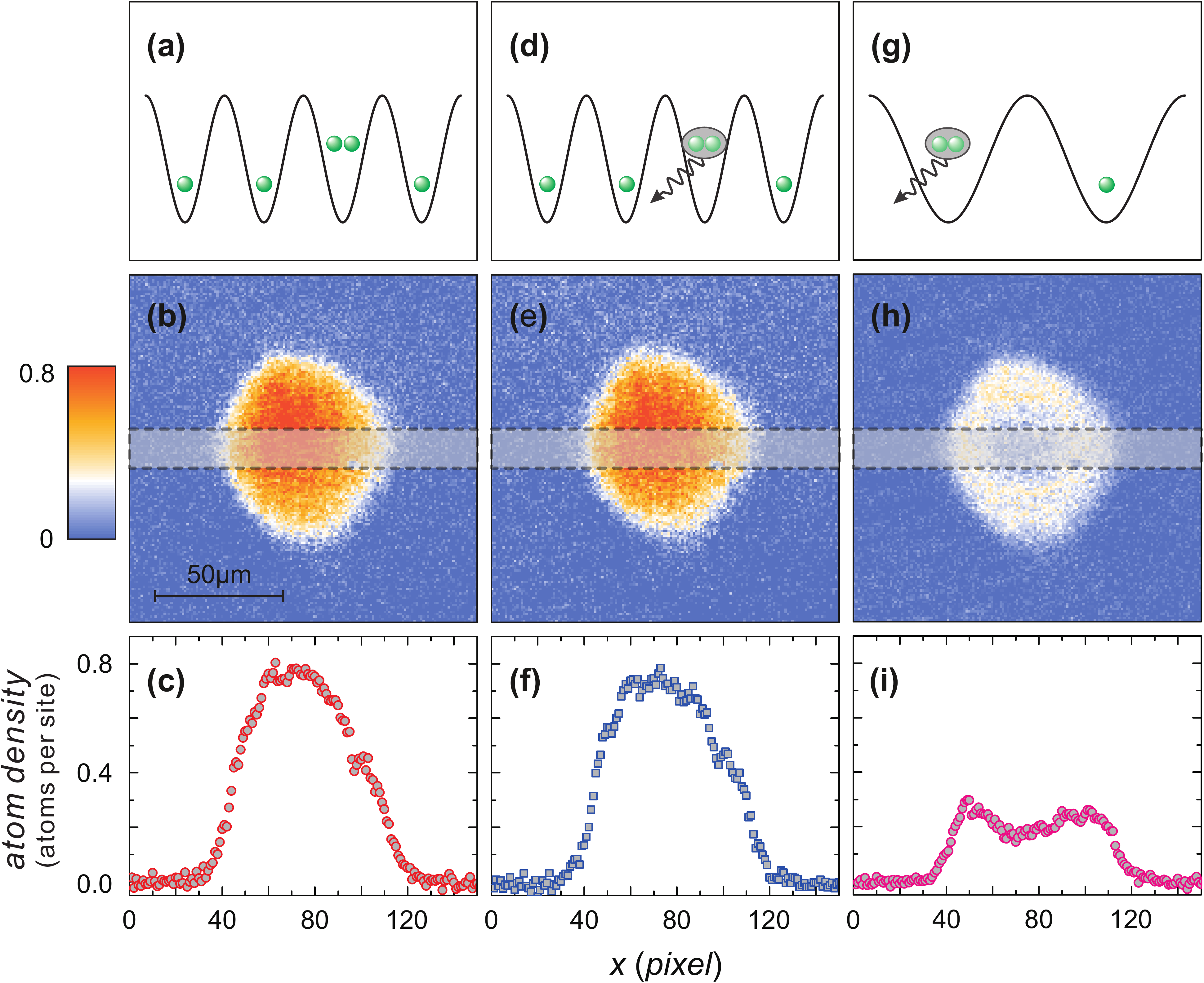}}%
\caption{{\bf Calibration of the filling parameters}: the schematics, in-situ distribution and the mean cross section along the X direction for the rectangular area (gray shading) with a width of  16 pixels (14.9 $\mu$m) for \textbf{(a,b,c)} the initial MI state; \textbf{(d,e,f)} the atom cloud after holding 500 ms in the short lattice; and \textbf{(g,h,i)} the atom cloud after the second filtering killing of 500 ms in the long lattice. No obvious loss is observed for the first filtering process, which indicates a low probability for filling with two atoms, while a clear loss is observed for the second filtering process, which corresponds to the filling of one atom in both sites of the DW.}
\end{figure}

We calibrate the hyperfine changing collisional loss in our experiment by comparing two loss processes: one when the lattice sites are mainly filled with one atom and the other one when they are doubly filled. To prepare a sample with high single occupancy, we first flip the left sites in the spin-dependent superlattice, remove them from the trap by an imaging pulse, and then transfer the remaining atoms into the long lattices (the average filling in the ROI here is about 0.4). Then we measure the collisional loss of the atoms in the trap by holding them inside the lattices (\hbox{$V_{xl}=35$} Er, \hbox{$V_{y}=60$} Er and pancake trap \hbox{$\omega_z=7$ }kHz) for different times. In this case, there is no obvious loss observed since there are no collisions for single atom, as shown in Fig.SI3. We intentionally prepare an atom cloud with high double occupancy by transferring the MI atoms into the long lattice. The loss process shows an exponential decay with a $1/e$ lifetime of 120(5) ms (Fig.S4), and all the double occupancies are removed from the lattice after holding in a deep lattice for 500 ms. One may expect a higher collision rate and faster loss process by tightening the confinement, however the collision rate  in our experiment is mainly limited by the confinement on the Z direction.

\begin{figure}[!t]
\centerline{\includegraphics[width=7.1 cm]{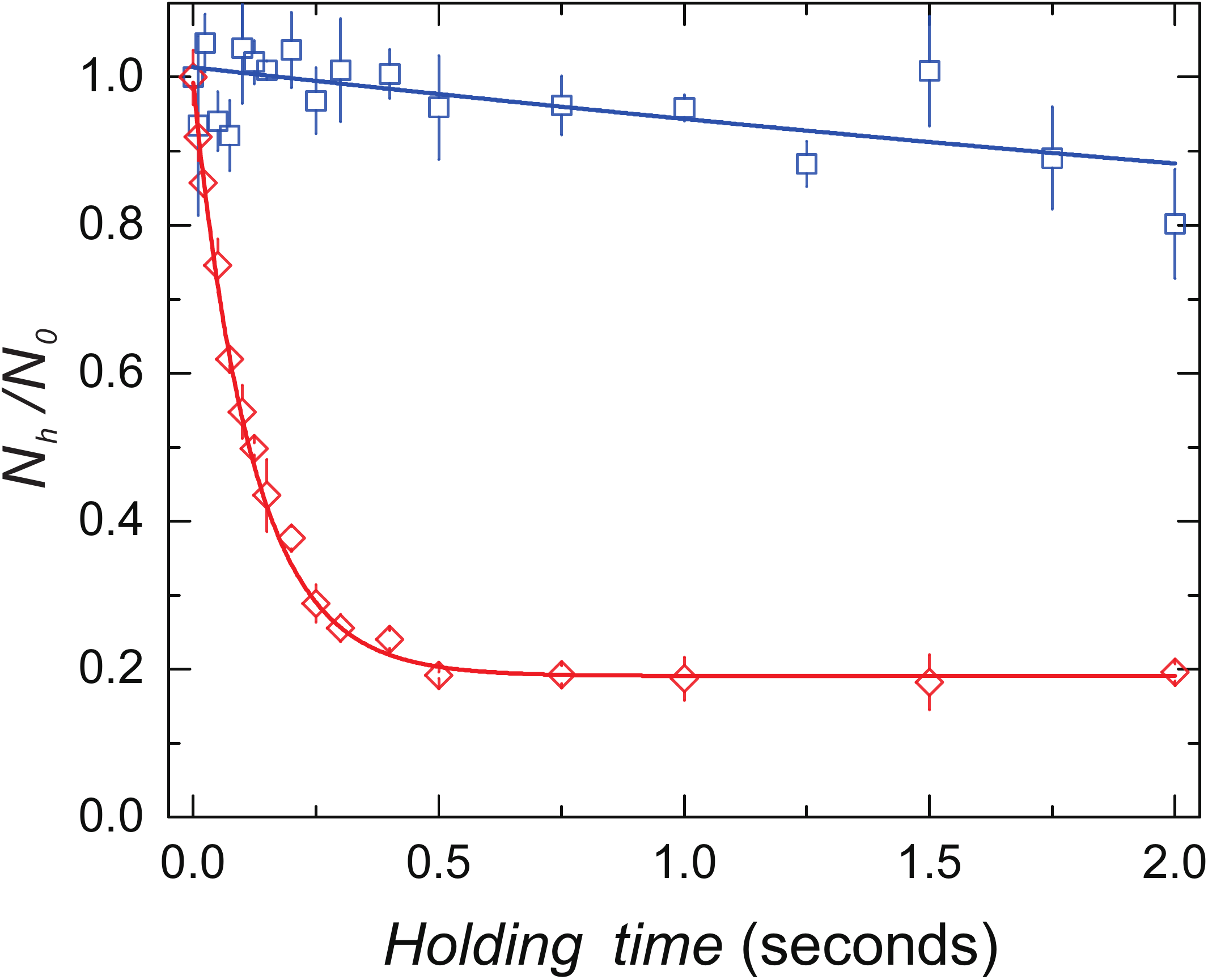}}
\caption{Measured collisional loss in the lattices by comparing the atoms in ROI with and without hyperfine spin changing collisions: a fast decay with time constant of 120 ms (red) is observed when some of the lattice site are filled with two atoms, while almost no decay is observed for the case of only one atom filled in lattice sites (blue), the fitting shows a $1/e$ lifetime of several seconds.}
\end{figure}

\subsection*{Calibration of the entanglement phase}
By applying a $\pi/2$-pulse on both atoms in the DW, the \hbox{$|\!\uparrow\rangle/|\!\downarrow\rangle$} basis are rotated to the \hbox{$|+\rangle/|-\rangle$} basis and an entangled state \hbox{$|\psi^{\prime}\rangle=(|\!\uparrow,\downarrow\rangle+e^{i\phi}|\!\downarrow,\uparrow\rangle) /\sqrt{2}$}  with entanglement phase $\phi$ is transferred to the following state as
\begin{equation}
|\psi^{\prime}\rangle\stackrel{\pi/2}{\Longrightarrow} \frac{1}{\sqrt{2}}(|+,-\rangle+e^{i\phi}|-,+\rangle)
= \frac{1-e^{i\phi}}{2\sqrt{2}} (|\!\uparrow,\downarrow\rangle - |\!\downarrow,\uparrow\rangle) + \frac{1+e^{i\phi}}{2\sqrt{2}}e^{i\pi/2} (|\!\uparrow,\uparrow\rangle + |\!\downarrow,\downarrow\rangle).
\end{equation}
For the triplet state, $\phi$ is 0, therefore \hbox{$|t\rangle \stackrel{\pi/2}{\Longrightarrow} (|\!\uparrow,\uparrow\rangle + |\!\downarrow,\downarrow\rangle)/\sqrt{2}$}. While for the singlet state, $\phi=\pi$, therefore $|s\rangle \stackrel{\pi/2}{\Longrightarrow} |s\rangle$. Following the two-stage filtering and imaging routine, the remaining atoms are proportional to the singlet fractions, i.e. \hbox{$\propto \!(1\!-\!\cos\phi)$}. The entanglement phase after a STO modulation is \hbox{$\phi(t_2)\!=\!\phi_0+\delta\!\cdot\! t_2$}, therefore the observed atom number for different STO times follows a cosine function in our experiment.

\subsection*{Calibration of the entanglement fidelity}
The fidelity of the two-atom entangled state generated in our experiment, represented by the density matrix $\rho$, with respect to the target state $|t\rangle$ can be calculated as \hbox{$F\!=\!\langle t|\rho|t\rangle$}. The lower boundary of the fidelity can be evaluated by following the analysis of Ref.[24] in the main text
\begin{equation}
\begin{split}
F \geq & (P_{\uparrow,\downarrow}+P_{\downarrow,\uparrow} - 2\sqrt{P_{\uparrow,\uparrow}\  P_{\downarrow,\downarrow}} + P_{+,+} +P_{-,-} - P_{+,-} - P_{-,+})\\
\geq& (P_{\uparrow,\downarrow}+P_{\downarrow,\uparrow} -  P_{\uparrow,\uparrow}- P_{\downarrow,\downarrow} + P_{+,+} +P_{-,-} - P_{+,-} - P_{-,+})\\
=&-(E_{zz}+E_{yy})/2
\end{split}
\end{equation}

\subsection*{Imaging system and calibration of atomic density}
The imaging objective with an effective numeric aperture of 0.48 contains an aspheric lens and a special designed glass plate. The glass plate is used to compensate the wavefront distortion of the glass cell. Another imaging lens with a focal length of 500 mm is used to direct the beam into the CCD camera. The amplification factor of the imaging system is $\textrm{M}\!=\!28.6$, and the pixel size on CCD camera corresponds to \hbox{$0.93\!\times\!0.93$} $\mu$m$^2$ in the real space at the atom position. With the absorption imaging of the atoms, the resolution is calibrated to be better than 2.3 $\mu$m.

To faithfully obtain the in situ densities of the atom cloud in the experiment, we apply the saturated absorption imaging technique[S2-S4]. A right circular polarized imaging beam with short duration of 10 $\mu$s and about 8 saturated intensities is applied along the Z direction to the two-dimensional atom cloud. The atomic density of the MI state in our experiment is lower than 1 atom per site, which corresponds to an optical density of \hbox{$od\!\approx\!2$}, thus the collective effect of the atom cloud is small. Furthermore, to avoid multi scattering processes in the highly compressed pancake-trap, we release the trap to lower down the 3D atomic density by switching off the lattices and pancake trap for 50 $\mu$s before applying the imaging pulse.
\begin{figure}[!t]
\centerline{\includegraphics[width=7.2cm]{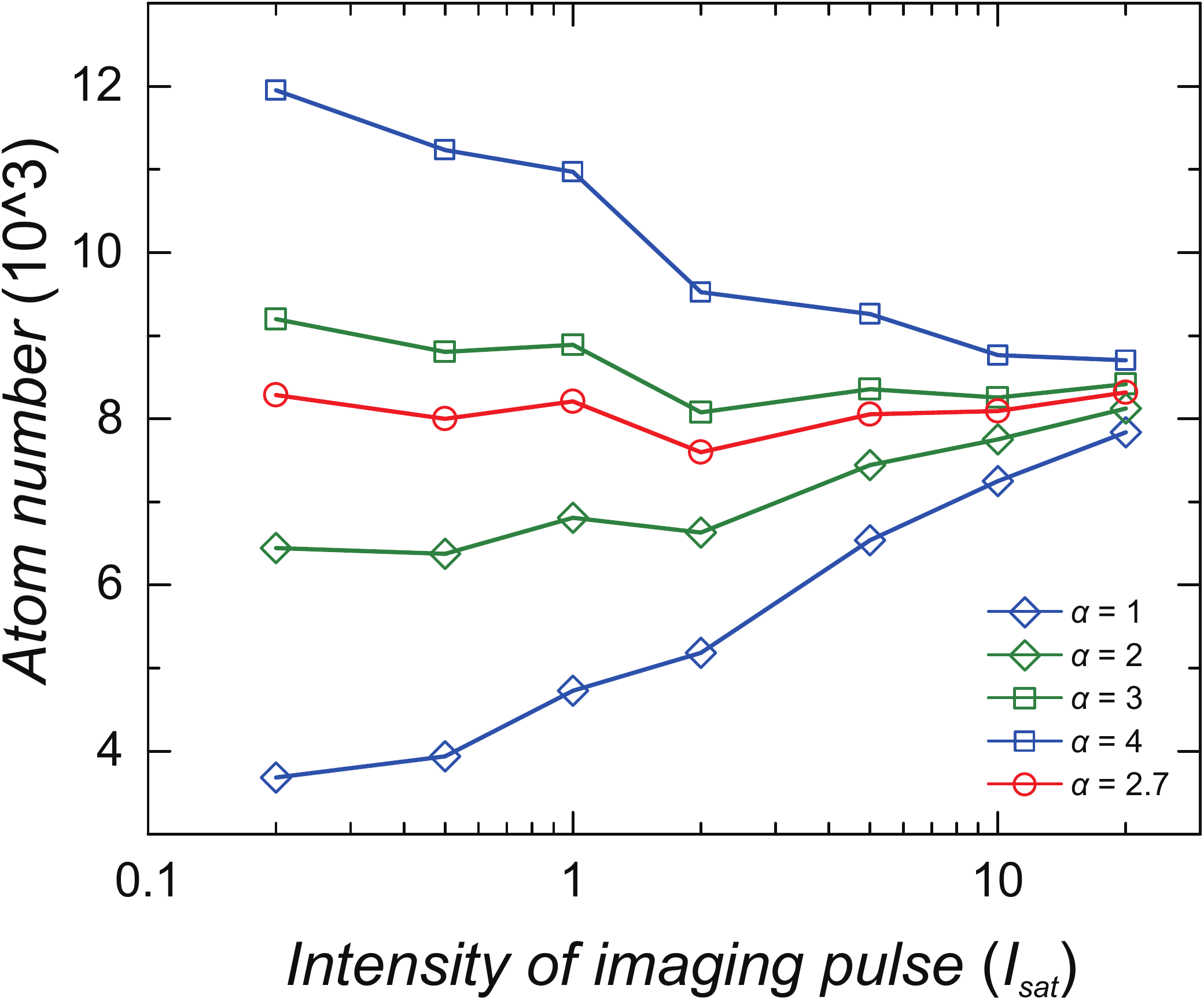}}
\caption{Calibration of the saturated absorption imaging with the modified Beer-Lambert law by applying different imaging intensities to a thermal atom cloud. When \hbox{$\alpha\!=\!2.7$}, almost the same atom number can be obtained for different imaging intensities.}
\end{figure}

We calibrate the saturated absorption imaging with the modified Beer-Lambert law$^{\rm S2}$
\begin{equation}
n(x,y)\sigma_{0} = -\alpha \ln \frac{I_t(x,y)}{I_i(x,y)}+ \frac{I_i(x,y)-I_t(x,y)}{I_{sat}}
\end{equation}
where $n(x,y)$ is the atomic density of the atom cloud, $\sigma_0$ is the scattering cross section for the cycling transition of \hbox{$5S_{1/2}|F\!=\!2\rangle\Leftrightarrow5P_{3/2}|F\!=\!3\rangle$} for circular polarized light, $\alpha$ is a scaling parameter, $I_i(x,y)$ is the intensity distribution of the imaging beam, $I_t(x,y)$ is the intensity distribution of the imaging beam after being absorbed by the atom cloud, and $I_{sat}$ is the saturated absorption intensity. We apply different intensities for the imaging pulse to a thermal atom cloud produced in the same condition. By optimizing the parameter $\alpha$, the same atom number is obtained for different pulses, as shown in Fig.S5. The calibration results \hbox{$\alpha\! = \!2.7(2)$} for our system. To suppress the large photon shot noise at high imaging intensities, a fringe removal algorithm[S5] is applied.

\subsection*{Reference}
\begin{itemize}
\item[[S1\!\!]]  I. H. Deutsch \& P. S. Jessen, {\it Physical Review A} {\bf 57}, 1972-1986 (1998).
\item[[S2\!\!]] G. Reinaudi, T. Lahaye, Z. Wang \& D. Gu¨¦ry-Odelin, {\it Optical Letters} {\bf32}, 3143-3145 (2007).
\item[[S3\!\!]] T. Yefsah, R. Desbuquois, L. Chomaz, K. G{\"u}nter \& J. Dalibard, {\it Physical Review Letters} {\bf 107}, 130401 (2011).
\item[[S4\!\!]] W. Muessel, H. Strobel, M. Joos, E. Nicklas, I. Stroescu, {\it et. al.}, {\it Applied Physics B} {\bf 113}, 69-73 (2013).
\item[[S5\!\!]] C.F. Ockeloen, A.F. Tauschinsky, R.J.C. Spreeuw and S. Whitlock, {\it Physical Review A} {\bf 82}, 061606 (2010).
\end{itemize}
\end{document}